\documentclass[fleqn,10pt]{wlscirep}
\usepackage[utf8]{inputenc}
\usepackage[T1]{fontenc}

\usepackage{algorithmicx}%
\usepackage{algpseudocode}%
\usepackage[ruled]{algorithm2e}
\usepackage{booktabs}
\usepackage{lmodern,babel,adjustbox,booktabs,multirow}
\usepackage{changepage} 
\usepackage[left, pagewise]{lineno}
\SetArgSty{textnormal}
\newcommand{\var}{\textit}
\usepackage{soul}

\title{Probabilistic metaplasticity for continual learning with memristors in spiking networks}

\author[1,*]{Fatima Tuz Zohora}
\author{Vedant Karia}
\author{Nicholas Soures}
\author{Dhireesha Kudithipudi}
\affil{University of Texas at San Antonio, Department of Electrical and Computer Engineering, San Antonio, TX 78249, USA}

\affil[*]{fatimatuz.zohora@my.utsa.edu}

\keywords{Metaplasticity, Continual learning, Memristor, Spiking neural network }

\begin{abstract}
Edge devices operating in dynamic environments critically need the ability to continually learn without catastrophic forgetting. The strict resource constraints in these devices pose a major challenge to achieve this, as continual learning entails memory and computational overhead.
Crossbar architectures using memristor devices offer energy efficiency through compute-in-memory and hold promise to address this issue. However, memristors often exhibit low precision and high variability in conductance modulation, rendering them unsuitable for continual learning solutions that require precise modulation of weight magnitude for consolidation. Current approaches fall short to address this challenge directly and rely on auxiliary high-precision memory, leading to frequent memory access, high memory overhead, and energy dissipation. In this research, we propose probabilistic metaplasticity, which consolidates weights by modulating their update \textit{probability} rather than magnitude. The proposed mechanism eliminates high-precision modification to weight magnitudes and, consequently, the need for auxiliary high-precision memory. We demonstrate the efficacy of the proposed mechanism by integrating probabilistic metaplasticity into a spiking network trained on an error threshold with
low-precision memristor weights. Evaluations of continual learning benchmarks show that probabilistic metaplasticity achieves
performance equivalent to state-of-the-art continual learning models with high-precision weights while consuming $\sim 67\%$ lower memory for additional parameters and up to $\sim 60 \times$ lower energy during parameter updates compared to an auxiliary memory-based solution. The proposed model shows potential for
energy-efficient continual learning with low-precision emerging devices.

\end{abstract}

\begin{document}

\flushbottom
\maketitle
%
%
\thispagestyle{empty}

\section*{Introduction}

Autonomous edge devices constitute a major application domain of artificial intelligence--ranging from indoor robots to autonomous drones/vehicles \cite{hayes2022online}. These devices often need on-device training for real-time adaptation to dynamic environments \cite{8778327}. However, training neural networks requires energy-intensive data movement and memory accesses, which can strain the energy budget of these resource-constrained platforms \cite{dally2022model}. The challenge exacerbates in real-world scenarios where data is non-stationary and does not maintain independent and identical distribution. Conventional artificial intelligence fails to learn effectively in this scenario and catastrophically forgets previously learned information \cite{FRENCH1999128}. The ability to learn and adapt to non-stationary data streams requires continual learning mechanisms, which entail additional memory and computational overhead, further straining the strict resource budget on edge devices \cite{verwimp2023continual}.
The constraints in autonomous edge applications present unique challenges, which lead to certain desirable criteria for continual learning solutions. First, the continual learning solution should be task-agnostic, i.e., it should be effective with no supervision or task-boundary awareness for autonomous applications. Second, edge devices have limited on-chip memory, which prohibits numerous iterative training over a large dataset. Hence, these devices should learn in an online scenario where data are encountered only once \cite{hayes2022online}. Finally, the memory and computational overhead of the continual learning mechanisms should be minimized considering the strict energy constraints of edge devices \cite{kudithipudi2023design}.

Recent research primarily focuses on alleviating catastrophic forgetting to achieve continual learning while overlooking the above-mentioned criteria. In our research, we strive to achieve continual learning in the online task-agnostic scenario while minimizing memory and computational overhead. We address this multifaceted challenge by taking inspiration from the biological brain, which shows the ability to learn continually throughout its lifetime with remarkable energy efficiency \cite{7155485}. Neuromorphic or brain-inspired approaches can help in two ways. First, we can take inspiration from the neural mechanisms deemed to help memory consolidation in the brain to address catastrophic forgetting
\cite{biological_underpinnings}. One such mechanism is metaplasticity or plasticity of plasticity \cite{abraham_meta}, whereby synapses modify their ability to change based on their history of activity. Recent research in continual learning abstracted this concept by detecting important weights and restricting their update in magnitude by modulating weight gradients. Metaplasticity-inspired weight consolidation mechanisms have shown promising performance in addressing continual learning in neural networks \cite{Benna2016, TACOS1, ewc, SI, pmlr-v80-kaplanis18a, laborieux2021synaptic}.
Second, we can harness the computational and structural principles of the brain to optimize the energy efficiency of on-device continual learning. This can translate to adopting spiking models that emulate the computational primitives of the brain with sparse binary encoding. This feature replaces energy-intensive floating-point operations with low-power event-driven computations \cite{SNN_benefits}. Structural primitives of the brain are emulated with architectures featuring co-located memory and processing with nanoscale emerging memristor devices \cite{brain_inspired_memristor}. In a crossbar arrangement, the conductance of these devices can emulate weights in a network and perform vector-matrix multiplication in a single timestep \cite{Xia2019}. Such architectures reduce data movement, enable parallel computations and can lead to up to two orders of magnitude greater energy efficiency compared to conventional architectures \cite{time}.

Although spiking models incorporated in memristor crossbar architectures are promising, they show non-ideal characteristics which can limit their suitability for continual learning mechanisms. The weight consolidation mechanism mentioned above is a prominent method of continual learning that relies on precise modulation of the weight magnitude. This is extremely challenging to achieve with memristors since these nanoscale devices show low precision ($\sim 3-6$ bits \cite{liehr2020impact, mem_64_level}) and inherent variability during conductance modulation. Recent research has circumvented this issue by incorporating auxiliary high-precision memory (usually implemented with Static Random Access Memory (SRAM)) alongside memristor crossbars \cite{comp_memory}. Binary memristor devices have been used to accelerate forward pass in continual learning tasks while training with backpropagation is carried out in high-precision memory \cite{mp_cont_learning}. Memristors with multi-level conductance have also been utilized for continual learning with metaplasticity, where auxiliary high-precision memory accumulates weight gradients, and memristor weights are updated when the accumulated gradient is equivalent to the device conductance resolution \cite{d2023synaptic}. These approaches have shown promising performance with low-precision memristor weights in continual learning benchmarks. However, these solutions rely on multi-epoch training and knowledge of task progression in order to either assign separate output neurons for different tasks or to modulate hyperparameters. These prerequisites are not suitable for autonomous applications. Moreover, high-precision memory with SRAMs lead to a large silicon footprint and idle standby leakage power (tens to hundreds of picoWatts per bit) \cite{Lu2024}. Finally, while crossbar architectures with auxiliary high-precision memory retain energy efficiency during inference, excessive memory access during training can potentially void the benefits of the compute-in-memory architecture.

In this research, we explore a novel weight consolidation mechanism suitable for low-precision memristor weights. Neuroscience literature has shown that binary metaplastic synapses characterized by probabilistic transitions help memory retention \cite{fusi2005cascade}.
Inspired by this, we propose the probabilistic metaplasticity model, where a weight's plasticity translates to its probability of update rather than its magnitude of update. The probability of a weight's update evolves during training such that important weights are assigned lower update probability to prevent catastrophic forgetting. Since this model does not need precise modulation in weight magnitude, the need for auxiliary high-precision memory is eliminated. Previous research explored probabilistic metaplasticity in the context of binary synapses in models of memory \cite{fusi2005cascade} and binarized networks that only support training a single-layer network \cite{metaplasticnet}. We generalize this concept to memristor weights with multi-level conductance incorporated in a multilayer spiking network trained based on an error threshold. 
We evaluate the proposed solution on image detection continual learning benchmarks considering single and multi-memristor weights. Evaluations show that low-precision memristor weights with probabilistic metaplasticity can achieve performance equivalent to state-of-the-art networks with high-precision weights in the challenging online task-agnostic continual learning scenario. Compared to an equivalent solution with high-precision auxiliary memory, probabilistic metaplasticity reduces memory overhead for additional parameters by $\sim 67\%$ while requiring up to $\sim 60 \times$ lower energy during the parameter update phase. The proposed model is promising for memory- and energy-efficient continual learning for autonomous edge applications.

\section*{Probabilistic metaplasticity with error threshold-based training}
Metaplasticity, or plasticity of plasticity, refers to the modification of a synapse or weight's ability/tendency to change based on previous activity \cite{abraham_meta}. It is deemed a key mechanism for memory consolidation in the brain \cite{meta_memory}. Fusi et al. proposed a cascade metaplasticity model, where binary synapses contain an additional discrete metaplastic state variable \cite{fusi2005cascade}. The metaplastic state of a synapse changes probabilistically in response to input stimuli. Synapses with higher metaplastic states are assigned a lower probability of change to signify greater stability.

\begin{figure*}[h!tb]
\centering

\includegraphics[width=1\linewidth]{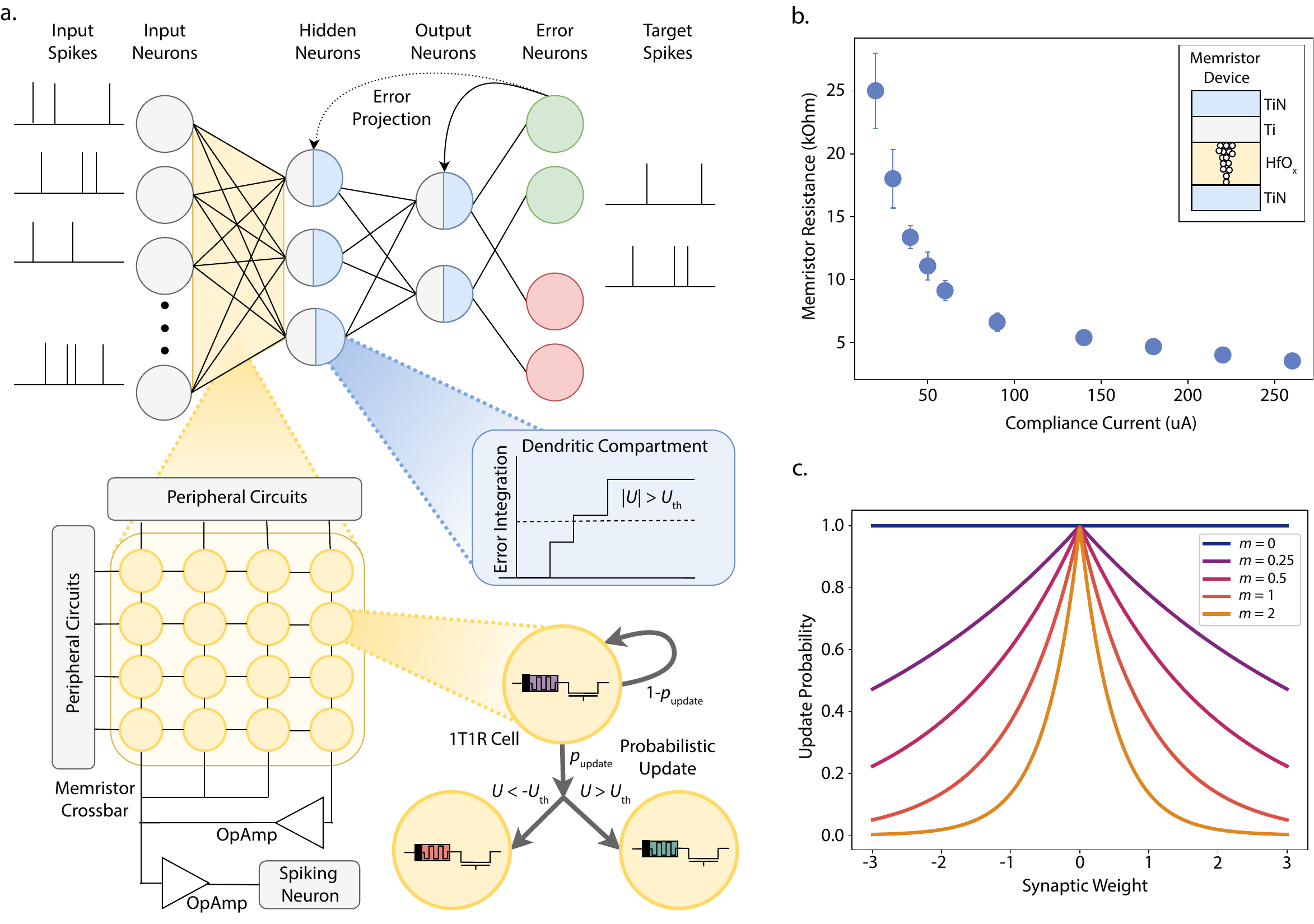}

\caption{\textbf{Probabilistic metaplasticity with error threshold-based training.} \textbf{a.} Spiking network trained on error threshold where the network weights are realized with 1T1R memristor crossbar array. The dendritic compartments in the hidden and output layer neurons integrate the error. When the error reaches the threshold $U_{\text{th}}$, the memristor weights are updated to the next higher (negative error) or lower (positive error) conductance level with probability $p_{\text{update}}$. \textbf{b.} Mean and standard deviation of resistance levels versus programming compliance current for the 1T1R memristor device (shown in inset) adopted in this work \cite{liehr2020impact}. \textbf{c.} Update probability of weights for different values of the metaplasticity coefficient $m$ and the weights $w$. $m$ is positively associated with the activity level of adjacent neurons. We see that weights with highly active adjacent neurons (high $m$) and weight magnitude (high $|w|$) lead to low update probability. }
\label{fig:concept_fig}
\end{figure*}

Similar to the discrete metaplastic states in the cascade model, memristors can be programmed to a limited number of distinguishable conductance levels (Figure \ref{fig:concept_fig}b). In our work, we consolidate memristor weights in a similar probabilistic manner. In a typical network, weights realized with memristors are updated deterministically by increasing or decreasing their conductances based on the error. In the proposed probabilistic metaplasticity model, the weights are updated probabilistically, where the important weights exhibit a lower probability. The importance of a weight is a function of its magnitude and the activity levels of adjacent neurons \cite{TACOS1}. High magnitude and connections to highly active pre- and post-synaptic neurons lead to a lower probability of updating or higher weight consolidation. This is captured in the metaplasticity function, which determines the update probability $p_{\text{update}}$ of a weight $w_{ij}$ connected between $i^{\text{th}}$ pre-synaptic and $j^{\text{th}}$ post-synaptic neurons - 

 \begin{equation}
    p_{\text{update}} = e^{-|m_{ij}w_{{ij}}|}.
    \label{eqn:meta_function}
\end{equation}

Here, $m_{ij}$ is a metaplasticity coefficient assigned to each weight. As the network learns, the metaplasticity coefficient evolves based on the activity trace of the adjacent neurons as -

\begin{equation}
    m_{ij}(t+1) =
\begin{cases}
    m_{ij}(t) + {\Delta}m, & \textrm{~if} \: X^{\text{tr}}_{i} \ge m^{\text{pre}}_{\text{th}} \& X^{\text{tr}}_{j} \ge m^{\text{post}}_{\text{th}} \\
    m_{ij}(t) & \textrm{~otherwise}
\end{cases}.
\label{eqn:m_evolution}
\end{equation}

We see that high activity in the adjacent neurons increases the metaplasticity coefficient of a weight.
Figure \ref{fig:concept_fig}c shows the update probability for different values of weights and metaplasticity coefficients. We see that weights with higher magnitude and metaplasticity coefficient exhibit lower update probability. Throughout training, a weight's magnitude and metaplasticity coefficient adapt in response to the activity of the network. This changes the update probability/plasticity of a weight, leading to a metaplastic effect to consolidate weights.

\begin{algorithm}[!t]
\DontPrintSemicolon

\SetKwComment{Comment}{$\triangleright$\  }{}
\SetKwComment{Commente}{$\#$ }{}

\SetKwFunction{Weight}{WeightUpdate}

    \For{$\textit{task} \in \mathcal{T}$}{

   \For{$\{x^t,y^t\} \in \{X^t,Y^t\}$}{
   
   \For{\var{t} $\in$ $T_{\text{sim}}$}{
   Forward Pass: $S^{\text{out}}(t)$ $\gets$ $f (S^{\text{in}}(t), W(t))$\;
   Random Error Feedback: $\tau_U \frac{\partial \mathrm{U}}{\partial t} =  E\mathrm{R}_U$\; 
   Update Neuron Trace: $\frac{d}{dt}X_{\text{tr}} = -\frac{X_{tr}}{\tau_{\text{tr}}} +S$ \;  
 \If{$|U_j|>U_{\text{th}}$}{
  \For {\var{i} $\in$ $\{X_i(t)==1\}$}{
   \If{$I_{\text{min}}<I_j<I_{\text{max}}$}{
   $\epsilon \sim \text{uniform}(0,1)$ \; 
   $p_{\text{update}} = e^{-|m_{ij}w_{{ij}}|}$\;
   \If {$\epsilon<p_{\text{update}}$}{
        \text{Update memristor weights}: $W$ $\gets$ $\text{Program} (R_{\text{mem}}, U)$\;
    }
    $U_j = 0 $\;
   }
   
   }

   }
   }
   Update metaplasticity coefficient: $m \gets m + {\Delta}m$ \;
   }
   }
\caption{Probabilistic metaplasticity with error threshold-based eRBP for continual learning in a spiking network. $\mathcal{T}$ denotes a set of sequential tasks. $S^{\text{in}}$ and $S^{\text{out}}$ are the input and output spike trains and $W$ denotes weights realized with memristors. $U$ is the error accumulated at the dendritic compartment of neurons. When |$U$| crosses the error threshold $U_{\text{th}}$, the update probability $p_{\text{update}}$ of eligible weights are calculated and compared with a random number to decide which weights should be updated. The memristor weights selected for update are programmed to the next higher or lower conductance level depending on the sign of the error. The function Program() refers to the operations required to update a memristor's conductance. Details of the eRBP algorithm can be found in methods.}
\label{alg:prob}
\end{algorithm}

We incorporate probabilistic metaplasticity into a multi-layer spiking network trained based on an error threshold (shown in Figure \ref{fig:concept_fig}a). The network is trained with event-driven random backpropagation (eRBP) where the network error is projected to neurons (details in Methods)\cite{neftci2017event}. Each neuron includes a dendritic compartment that accumulates the error. In eRBP, a weight is considered eligible for update if the pre-synaptic neuron has an active spike and the post-synaptic current is within a range. When these criteria are met, the weights are updated in proportion to the accumulated error. However, low-precision weights cannot be updated with arbitrary resolution. We resolve this issue by updating the weights when the accumulated error reaches a threshold. The threshold corresponds to the magnitude of error that warrants a change in the weights equivalent to their resolution. This update criteria is similar to the gradient accumulation approach where high-precision memory accumulates the gradients, and weights are updated when their gradients reach a certain threshold \cite{comp_memory}. The distinctive factor in our approach is that the weight update decision depends on the local error estimated at the neurons instead of gradients accumulated for every weight. This avoids gradient computation and storage and leads to lower memory overhead and accesses. 
Spiking networks trained with error-triggered learning can effectively learn a single task \cite{error-triggered}. We combine probabilistic metaplasticity with error threshold-based training to continually learn sequential tasks. Eligible weights are probabilistically updated, where the weight update involves programming the memristor device to its adjacent higher or lower conductance level, depending on the sign of the error. Important weights exhibit low update probability and hence retain their state and help prevent catastrophic forgetting. The detailed learning algorithm is described in Algorithm \ref{alg:prob}.

\section*{Results}
\subsection*{Continual learning with probabilistic metaplasticity }
We evaluate the proposed probabilistic metaplasticity model considering weights realized with a Hafnium Oxide-based memristor device in the IBM 65nm 10LPe CMOS/Memristor process \cite{liehr2020impact}. The device can be programmed to ten distinguishable low resistance levels which are used to map the weights (shown in Figure \ref{fig:concept_fig}b). The average resolution of conductance is $\sim$ 3 bits considering the average conductance at different levels. To realize both positive and negative weights, memristor crossbars are assigned a bias column whose contributions to the output are subtracted from that of the weights with inverting and summing amplifiers \cite{bias_col_paper}. If $g_\text{p}$ is the conductance of the memristor weight, $g_\text{b}$ is the conductance of the bias memristor, then the synaptic weight can be expressed with Equation \ref{eq:synaptic weight} where $g_\text{f}$ is the scaling factor implemented with the conductance of the feedback resistor in the inverting amplifier.

\begin{equation}
    w = \frac{1}{g_\text{f}}(g_\text{p}-g_\text{b})
    \label{eq:synaptic weight}
\end{equation}

We evaluate a spiking network with memristor weights on image detection continual learning benchmarks. The network is trained sequentially on tasks without any knowledge of the identity of the task or its boundaries during training or inference. This scenario is known as domain incremental learning (Domain-IL), which presents two major challenges \cite{vandeVen2022, Hsu18_EvalCL}. First, if the task boundary is known, the network can expect a change in data distribution and take measures to reduce catastrophic forgetting. This is not possible in domain-IL. Second, all tasks share the same output layer, which leads to increased interference and catastrophic forgetting. We also consider an online scenario where the network sees each sample only once (the network is trained on each task for one epoch). 

\begin{figure*}[h!tb]
\centering

\includegraphics[width=1\linewidth]{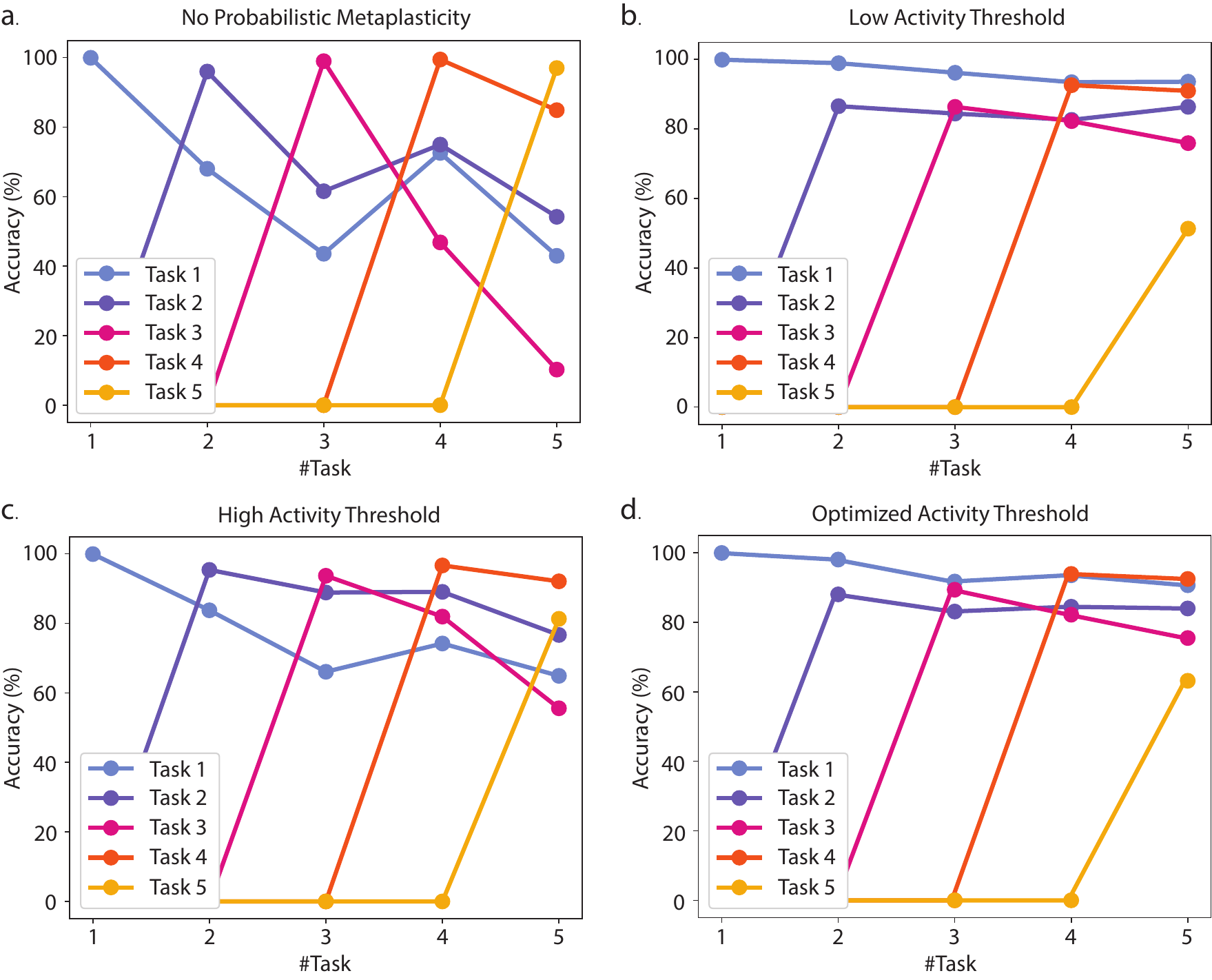}

\caption{\textbf{Evolution of task accuracies with sequential training on the split-MNIST benchmark.} The x-axis shows the latest task the network has learned. We evaluate the performance of task $n$ only after the network encounters it and denote the accuracies as 0 before that. \textbf{a.} With no probabilistic metaplasticity, the network learns the current task well, but forgets the initial tasks after sequentially learning multiple tasks. \textbf{b.} Probabilistic metaplasticity with low activity threshold leads to high rigidity in the network, so it remembers previous tasks but cannot learn the last task. \textbf{c.} High activity threshold can lead to loss in previous task accuracy while the network remains plastic to learn the new tasks. \textbf{d.} Optimized activity threshold balances plasticity and rigidity such that the network maintains high initial task accuracies while maintaining the ability to learn new tasks.}
\label{fig:acc_evolution}
\end{figure*}

We first evaluate probabilistic metaplasticity on the split-MNIST benchmark, which consists of five sequential image classification tasks, each composed of two classes from the MNIST dataset \cite{MNIST_dataset}. We conduct network-level simulations in Python considering a spiking network with 200 hidden neurons and 2 output neurons. The simulation emulates in situ learning with memristor weights and captures the quantized nonuniform distribution and the associated variance of memristor conductance during weight updates. Figure \ref{fig:acc_evolution} shows the evolution of the network performance in each of the benchmark tasks as the network trains on them sequentially. When trained with error threshold-based eRBP with no probabilistic metaplasticity, we see that the network learns each task it encounters with high accuracy (Figure \ref{fig:acc_evolution}a). However, after training on all tasks, it retains high accuracy only on the recently encountered tasks, and the earlier task accuracies drop to random guess level. 

Probabilistic metaplasticity consolidates important weights to alleviate such catastrophic forgetting. The degree of consolidation depends on the activity threshold of pre-synaptic and post-synaptic neurons ($m^{\text{pre}}_{\text{th}}$ and $m^{\text{th}}_{\text{post}}$ in Equation \ref{eqn:m_evolution}). A low activity threshold leads to high weight consolidation, which helps retain high accuracy in earlier tasks, as shown in Figure \ref{fig:acc_evolution}b. However, the network cannot adapt when later tasks are encountered due to the lower plasticity of the weights, leading to poor performance. A high activity threshold leads to the opposite effect, where the network can learn later tasks better with a compromise in earlier task accuracies (Figure \ref{fig:acc_evolution}c). Optimizing the activity threshold balances the stability-plasticity trade-off to retain high accuracy in earlier tasks with minimal degradation in the ability to learn later tasks (Figure \ref{fig:acc_evolution}d).

We further evaluate probabilistic metaplasticity on the split-Fashion MNIST benchmark, which also consists of five sequential image classification tasks, each with two classes \cite{xiao2017fashion}. We compare the performance of probabilistic metaplasticity with state-of-the-art models which also consolidate important weights or regularize weight updates to achieve continual learning. These approaches differ in the criteria that determine weight importance. ``Synaptic Intelligence'' (SI) \cite{SI} measures weight importance based on the contribution of weights to improve network performance, whereas ``Memory Aware Synapses'' (MAS) \cite{Aljundi_MAS} measures it through the sensitivity of the network's output function to changes in weights. ``Stochastic Synapses'' (SS) \cite{SS} adopts weights which transmit information probabilistically and weights with high transmission probability are consolidated. ``Bayesian Gradient Descent'' (BGD) \cite{zeno2018task} is a Bayesian continual learning mechanism that assesses the importance of parameters based on the uncertainty associated with their distribution. ``Learning without forgetting'' (LwF) \cite{li2017learning} preserves the response of the network to previous tasks through knowledge distillation. ``TACOS'' \cite{TACOS1} introduces activity-dependent metaplasticity where weights are consolidated based on their magnitude and the activity level of adjacent neurons. Except for TACOS, which was evaluated in a spiking network, the rest of the models were evaluated considering multi-layer perceptrons. Weight consolidation models typically require additional memory to store information for consolidating weights, which can be implemented with on-chip memory in hardware. In this work, we model this memory as on-chip SRAM during the energy analysis in the following subsection. Table \ref{tab:cont_acc} shows the memory required by each continual learning model for parameters, excluding weights, considering the same network architecture used to evaluate probabilistic metaplasticity (see Supplementary note S1 for details). We exclude the network weights in the memory overhead computation since the aforementioned continual learning models were evaluated considering full-precision weights, whereas the proposed mechanism realizes weights with memristor devices. We also evaluate a spiking neural network with full-precision weights trained with eRBP as a baseline.

Table \ref{tab:cont_acc} lists the mean accuracy across tasks for the different models after sequentially training on multiple tasks. To observe the effect of weight resolutions on the performance of probabilistic metaplasticity, we consider both single- and multi-memristor weights\cite{boybat2018neuromorphic}. Multi-memristor weights are configured by connecting $n_{\text{mem}}$ number of memristors in parallel. Figure \ref{fig:concept_fig}a shows a 1T1R crossbar with multi-memristor weights consisting of three 1T1R devices. During training, only one memristor arbitrated by a global counter is updated to reduce training overhead. The parallel combination of devices increases the dynamic range of the weight conductance by a factor of $n_{\text{mem}}$ and increases the weight resolution. 
 We note an improvement in the mean accuracy across tasks with increased weight resolution, with the best performance when $n_{\text{mem}}$ = 7, leading to a mean weight resolution $\sim 6$-bit. \textit{Results show that with probabilistic metaplasticity, a spiking network with noisy low-precision multi-memristor weights achieves performance equivalent to state-of-the-art networks with full-precision weights in both benchmark tasks.} For reference, we consider a spiking network that consolidates weights with activity-dependent metaplasticity (details in Methods) \cite{TACOS1}. Since this weight consolidation mechanism requires precise weight updates, we accumulate the weight gradients in auxiliary memory, which is reset after every sample. This model is evaluated with multi-memristor weights ($n_{\text{mem}}$ = 7) considering the same network architecture as probabilistic metaplasticity. We see that activity-dependent metaplasticity leads to performance similar to probabilistic metaplasticity in benchmark tasks. However, the latter requires $\sim 67\%$ lower memory overhead. We extend the performance analysis by evaluating probabilistic metaplasticity on the split-CIFAR-10 benchmark \cite{cifar10}, which consists of five sequential image detection tasks. It is a more complex dataset compared to the previous benchmarks, containing larger images, more complex image types and colored images instead of grayscale. 
 To address this complexity, we consider a ResNet-18 backbone pre-trained on ImageNet to process the images and use the extracted features as inputs to the spiking networks \cite{NEO, vandeVen2022}. Probabilistic metaplasticity was evaluated considering a spiking network with 200 hidden neurons and multi-memristor weights ($n_{mem}$ =7). We also evaluate a baseline spiking network with multi-memristor weights and a spiking network trained with TACOS. Table \ref{tab:CIFAR-10} lists the accuracies of individual tasks and mean accuracy across tasks. We see that although probabilistic metaplasticity adopts low-precision multi-memristor weights, it demonstrates performance equivalent to TACOS, which adopts full-precision weights. 
 
\begin{figure*}[t]
\centering
\includegraphics[width=0.5\linewidth]{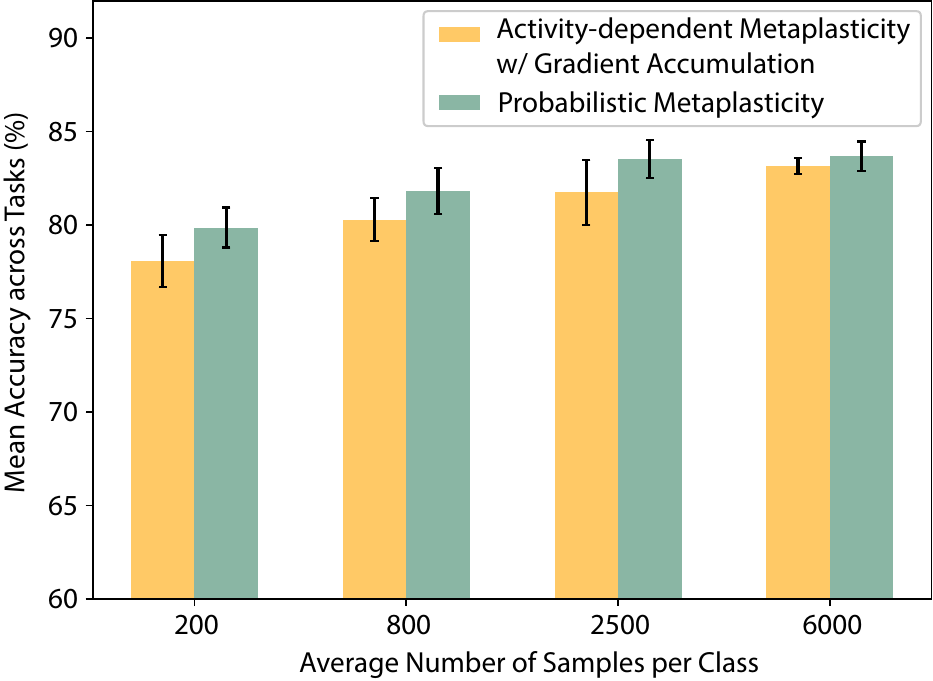}
\caption{\textbf{Effect of number of training samples on continual learning.} The mean accuracy across tasks shows degradation as the average number of samples per class reduces.}

\label{fig:sample_size}
\end{figure*}

 We further analyze the robustness of the proposed probabilistic metaplasticity model by evaluating its performance on a smaller set of training data. An autonomous agent learning in an online setting may need to learn from fewer samples. To explore this scenario, we trained and optimized the model on a subset of the split-MNIST dataset and tested it on the complete test set. We observed some degradation in the mean accuracy across tasks with reduced dataset size (shown in Figure \ref{fig:sample_size}. The degradation may stem from the probabilistic consolidation mechanism's reliance on the law of large numbers. However, we observed a similar degradation when training the network with activity-dependent metaplasticity. This suggests that the decline in performance could also be attributed to the general performance degradation in machine learning models when the number of training samples is insufficient.

Finally, we perform two controlled experiments to show the contribution of probabilistic metaplasticity for continual learning. First, we study the contribution of consolidating weights with high magnitude and high adjacent neural activity. For this study, we explore random consolidation in which the update probabilities of the weights are calculated with equation \ref{eqn:meta_function}, but the results are randomly shuffled among the weights. Second, we study the contribution of probabilistic metaplasticity in contrast to probabilistic plasticity. The controlled experiment evaluates this by updating all weights with the same update probability. We also reduce the weight update probability or plasticity by a factor with each incoming task to help retain previous knowledge. 
Table \ref{tab:control_exp} shows the results of the controlled experiments on the split-MNIST benchmark. We carried out the experiments considering multi-memristor weights ($n_{\text{mem}}$ = 7) in the same spiking network as in the previous experiments. The baseline network is trained based on an error threshold with no additional mechanism. The table lists the accuracies of each task in the benchmark and the mean accuracy across tasks after sequential training. We see that random consolidation fails to learn continually and suffers from catastrophic forgetting. The performance of the decaying probabilistic plasticity varies with the update probability reduction factor. A high reduction factor leads to consolidation in weights, which retains high accuracies in earlier tasks. 
However, the network fails to learn the last task. A low reduction factor shows the opposite characteristic, where the network catastrophically forgets the earlier tasks. Controlled experiments on the split-Fashion MNIST benchmark show the same response (Supplementary note S2 and Supplementary Table S2).
Contrary to the experiments, we see that probabilistic metaplasticity retains high accuracy in earlier tasks with much lower degradation in the accuracy of the last task compared to decaying probabilistic plasticity. Since the proposed mechanism consolidates weights selectively based on their magnitude and adjacent neural activity, 
it retains high plasticity in weights assigned low importance, leading to a better balance of stability and plasticity.

In summary,  probabilistic metaplasticity demonstrates continual learning with low-precision memristor weights in the challenging online task-agnostic continual learning scenario, which has not been addressed before in the literature. It achieves state-of-the-art performance with low memory overhead, which is highly suitable for edge applications.

\begin{figure*}[t]
\centering
\includegraphics[width=1\linewidth]{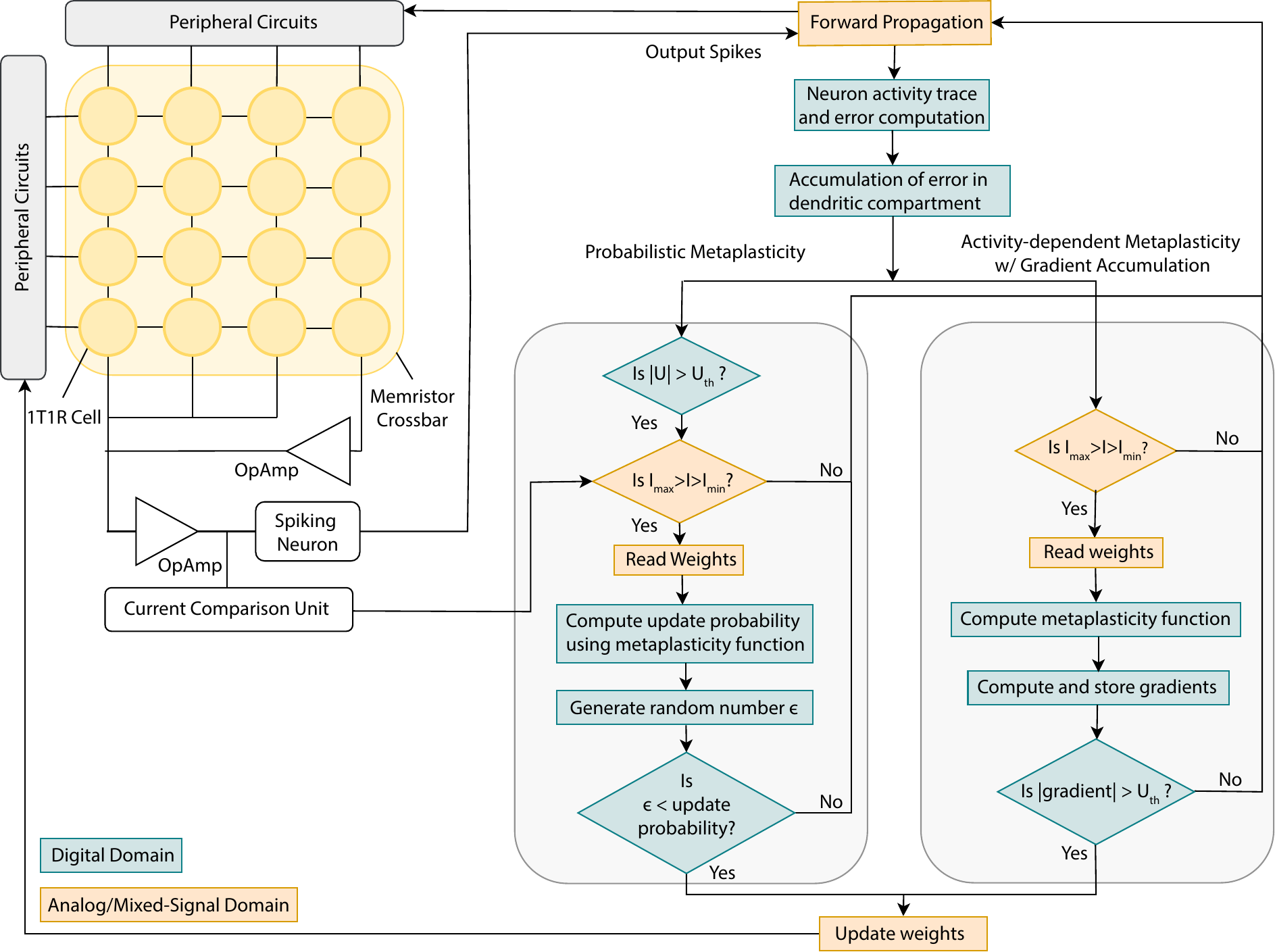}
\caption{\textbf{Mixed-signal architecture and computational flow for the continual learning mechanisms}. We consider a mixed-signal architecture where memristor crossbars carry out the forward pass of the spiking model and the error computation, arithmetic and logical operations are carried out in the digital domain. The figure shows the computational flow for continual learning with probabilistic metaplasticity (left) and activity-dependent metaplasticity with gradient accumulation (right).}
\label{fig:energy_flow}
\end{figure*}

\subsection*{Energy analysis}
We evaluate energy consumption for continual learning with probabilistic metaplasticity by analyzing the computations and memory accesses required to train on the benchmark tasks. We then determine the energy needed to train on a sample by analyzing the average number of operations and the individual operation energy requirements (details in Supplementary note S3). We also perform the analysis for activity-dependent metaplasticity with gradient accumulation as a reference. Figure \ref{fig:energy_flow} shows the architecture and computational flow for the analysis. We focus on the energy consumption during the parameter update phase (the operations in the gray box in Figure \ref{fig:energy_flow}) since the operations incurred during the forward pass and error computations are the same in both approaches.
We analyze the activities of a spiking network with 200 hidden neurons and multi-memristor weights ($n_{\text{mem}}$ = 7). The mixed-signal and digital components for different network operations were designed in the IBM 65nm 10LPe process. We designed a $16\times16$ 1T1R crossbar array with peripherals to estimate the energy to read and write memristor weights. The metaplasticity coefficients for both mechanisms and the gradients for the activity-dependent metaplasticity mechanism were modeled as on-chip SRAM with 16-bit and 32-bit bus widths, respectively (more details in Supplementary note S3). 

Figure \ref{fig:en_breakdown}a shows the energy consumption per sample during the parameter update phase and its breakdown. 
 We highlight the energy consumption due to memory access with memristor and SRAM read/write and subsume the energy required for the rest of the operations (post-synaptic current comparison, random number generation, error computation, etc.) in the "Other operations" category. We observe that probabilistic metaplasticity leads to a significant reduction in energy. The reason behind this is mainly twofold. First, the operations required for updating weights with probabilistic metaplasticity require lower energy compared to the gradient accumulation-based approach. Each time a weight is eligible for update, activity-dependent metaplasticity with gradient accumulation requires computing the gradient and the metaplasticity function. The latter requires reading the metaplasticity coefficient from SRAM and reading memristor weights. The computed gradient is multiplied by the output of the metaplasticity function and accumulated, which requires an SRAM read and write. On the other hand, probabilistic metaplasticity avoids memory access due to gradient accumulation. It also updates weights based on the comparison of a random number with the update probability (computed with the metaplasticity function), which requires much less energy compared to multiplication. Second, in the gradient accumulation-based approach, the weights are updated based on the criteria set in eRBP, while in probabilistic metaplasticity, the eligibility for weight update includes an additional condition of the error threshold. This leads to much fewer weights eligible for update in the latter (up to three orders of magnitude reduction on average), which consequently leads to less frequent execution of the operations needed for weight update. In summary, gradient accumulation leads to a higher fraction of weights eligible for update, and each eligible weight requires a higher number of memory accesses and computations for gradient computation, leading to higher energy consumption. We observe up to $\sim 60 \times$ reduction in total energy during parameter updates with probabilistic metaplasticity compared to activity-dependent metaplasticity with gradient accumulation.

\subsection*{Memory optimization with parameter sharing}
Energy analysis shows that SRAM read/write dominates energy consumption in both continual learning approaches. In the probabilistic metaplasticity model, the SRAM read/write operations account for reading and updating the metaplasticity coefficients assigned to the weights. Since a metaplasticity coefficient is assigned to each weight, the area overhead due to this memory can be substantial. Here, we optimize this memory overhead by sharing metaplasticity coefficients among weights. 
We  explore sharing the metaplasticity coefficients based on neural connectivity where weights connected to the same post-synaptic neuron share a metaplasticity coefficient (neuron-shared $m$). The metaplasticity coefficient associated with the weights connected to the $j^{\text{th}}$ post-synaptic neuron evolves as -
\begin{equation}
    m_{j}(t+1) =
\begin{cases}
    m_{j}(t) + {\Delta}m, & \textrm{~if} \:  X^{\text{tr}}_{j} \ge m^{\text{post}}_{\text{th}} \\
    m_{j}(t), & \textrm{~otherwise}
\end{cases},
\label{m_neqn}
\end{equation}

\begin{figure*}[t]
\centering

\includegraphics[width=\linewidth]{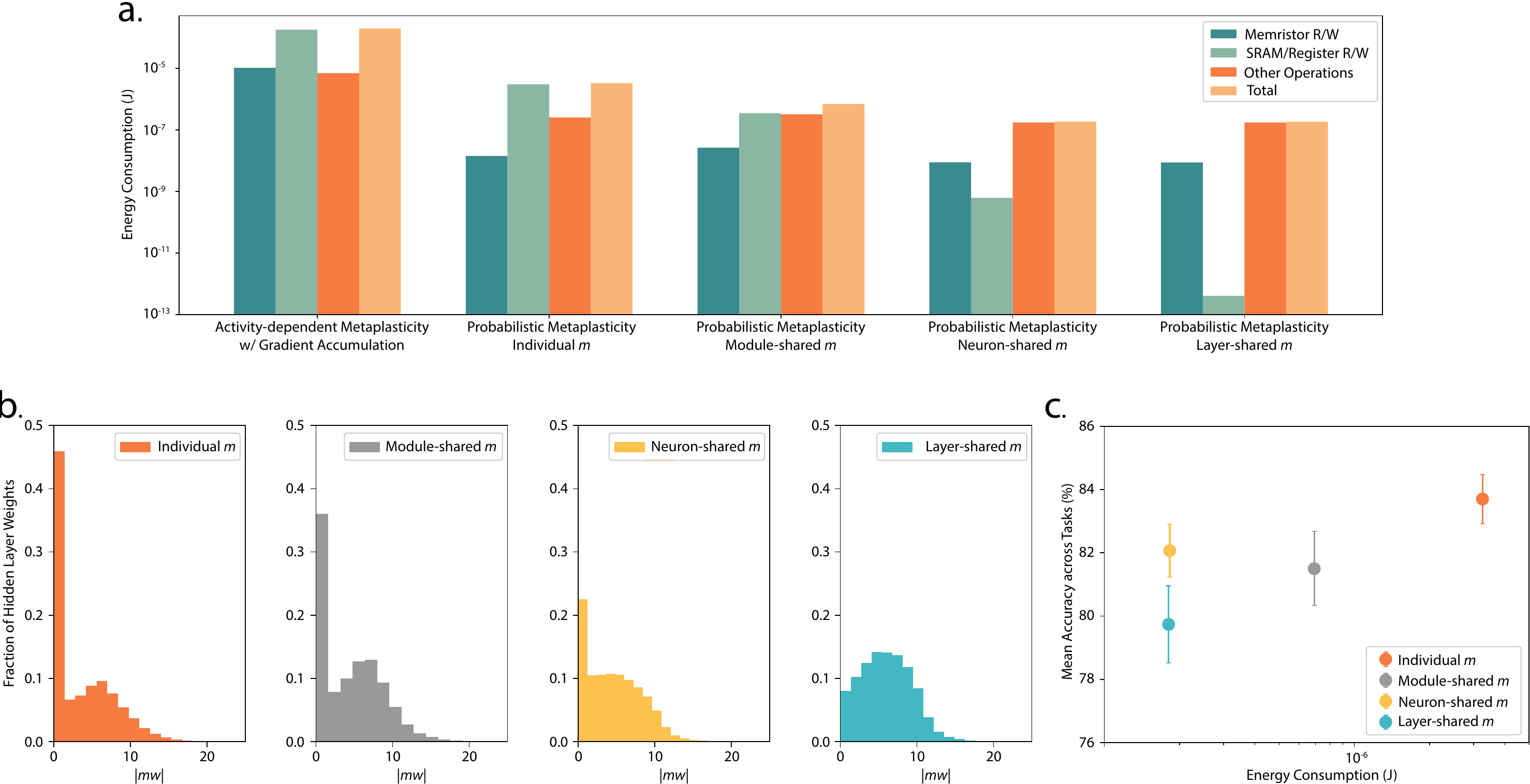}
\caption{\textbf{Energy consumption and stability-plasticity tradeoff.} \textbf{a.} Energy consumption per sample and its breakdown during the parameter update phase for activity-dependent metaplasticity with gradient accumulation and probabilistic metaplasticity with individual and shared metaplasicity coefficients.  \textbf{b.} Distribution of the product of metaplasticity coefficient $m$ and weight magnitude $|w|$ in the hidden layer with probabilistic metaplasticity. We see that as more weights share $m$, lower fraction of weights assume low $|mw|$ value. This indicates reduced plasticity in the network. \textbf{c.} Mean accuracy across tasks vs. the energy consumption per sample for the split-MNIST task. We observe a tradeoff between continual learning performance and energy consumption as the metaplasticity coefficients are shared across weights.}
\label{fig:en_breakdown}
\end{figure*}

where $X^{\text{tr}}_j$ is the activity trace of the $j^{\text{th}}$ post-synaptic neuron.
This configuration reduces the memory overhead substantially since only one coefficient is required per post-synaptic neuron in a layer.
To further optimize the memory overhead, we also consider assigning all weights in the same layer the same metaplasticity coefficient (layer-shared $m$). In this configuration, the metaplasticity coefficient is updated based on the average activity trace of all post-synaptic neurons in the layer -

\begin{equation}
    m(t+1) =
\begin{cases}
    m(t) + {\Delta}m, & \textrm{~if} \:  \frac {(\sum_{j=1}^{n}X^{\text{tr}}_{j})}{n} \ge m^{\text{layer}}_{\text{th}} \\
    m(t), & \textrm{~otherwise}
\end{cases},
\label{m_leqn}
\end{equation}
where $n$ is the number of post-synaptic neurons in a layer.

We evaluate the probabilistic metaplasticity model with shared metaplasticity coefficients on a spiking network with 200 hidden neurons and multi-memristor weights ($n_{\text{mem}}$ = 7). Table \ref{tab:neu_table} lists the evaluation results on the benchmark tasks. We see that probabilistic metaplasticity with neuron-shared and layer-shared $m$ performs on par or better compared to state-of-the-art models with weight consolidation based on neural activity \cite{NAI, NEO}. Furthermore, sharing metaplasticity coefficients leads to a substantial decrease in memory overhead and energy consumption during training. We model the metaplasticity coefficients neuron-shared and layer-shared $m$ with on-chip SRAM and registers, respectively, during energy analysis. Figure \ref{fig:en_breakdown}a shows, in both neuron-shared and layer-shared configurations, energy consumption is no longer dominated by SRAM read/write, rather most of the energy is spent in operations such as assessing the weight update criteria and memristor read/write. Hence, although layer-shared $m$ drastically reduces memory overhead and leads to three orders of magnitude reduction in energy dissipation to read/write the metaplasticity coefficients compared to neuron-shared $m$, in both cases we observe $\sim 18 \times$ reduction in total energy consumption during the parameter update phase.

The reduction in energy consumption, however, is accompanied by a degradation in performance compared to the probabilistic metaplasticity model with individual metaplasticity coefficients. This can be attributed to the limited flexibility of the network to consolidate weights, which leads to a poorer balance of the stability-plasticity dilemma. In the split-MNIST task, sharing metaplasticity coefficients leads to reduced plasticity after sequential training. This can be observed from the update probability of the weights, which is inversely related to the product of the metaplasticity coefficient and the weight magnitude ($|mw|$). The distribution of $|mw|$ in hidden weights (Figure \ref{fig:en_breakdown}b) shows that shared metaplasticity coefficients lead to a lower fraction of weights with $|mw|$ close to zero, indicating reduced plasticity. This reduction is more pronounced in layer-shared $m$. Individual task accuracies after sequential training also show that shared coefficients result in poorer accuracies in later tasks (Supplementary Table S5). This demonstrates a trade-off in the memory overhead and the network's ability to balance stability and plasticity.

Despite the reduction in plasticity, the substantial gain in energy consumption incentivizes shared metaplasticity coefficients in extreme-edge applications. The trade-off between memory overhead and the stability-plasticity balance can be further optimized by sharing the metaplasticity coefficients among blocks of weights within a layer. 
We investigated a modular sharing of metaplasticity coefficients (module-shared $m$), where a block of weights, connecting a post-synaptic neuron to a set of pre-synaptic neurons, shares a metaplasticity coefficient. The pre-synaptic neurons can be selected based on different criteria such as the sign or magnitude of incoming weights. Here, we group pre-synaptic neurons based on proximity where adjacent pre-synaptic neurons are grouped together. This configuration is well-suited for hardware implementation, as the activity of adjacent neurons can be computed locally, minimizing communication overhead.
The coefficient $m_{kj}$, assigned to module $k$, consisting of post-synaptic neuron j and $n_{\text{block}}$ number of adjacent pre-synaptic neurons, are updated based on the post-synaptic activity trace and the average of the traces of the pre-synaptic neurons in the module as follows -

\begin{equation}
    m_{kj}(t+1) =
\begin{cases}
    m_{kj}(t) + {\Delta}m, & \textrm{~if} \: \frac {(\sum_{i=1}^{n_{\text{block}}}X^{\text{tr}}_{i})}{n_{\text{block}}} \ge m^{\text{pre}}_{\text{th}} \& X^{\text{tr}}_{j} \ge m^{\text{post}}_{\text{th}} \\
    m_{kj}(t) & \textrm{~otherwise}
\end{cases}.
\label{eqn:m_block}
\end{equation}

We evaluate the performance of probabilistic metaplasticity with module-shared $m$ in the split-MNIST task considering sharing $m$ among 8 and 4 adjacent weights in the hidden and the output layer, respectively. Figure \ref{fig:en_breakdown}c shows the mean accuracy across tasks for different configurations of shared $m$ vs. the energy consumption per sample during the parameter update phase. We see that while the performance of module-shared $m$ is similar to that of neuron-shared $m$, the former shows greater plasticity compared to neuron-shared $m$ (Figure \ref{fig:en_breakdown}b). It also reduces memory overhead by $\sim 87.5\%$ and reduces energy consumption by $\sim 4.7\times$ during the parameter update phase compared to individual $m$. 
This configuration can be a promising approach to optimize the trade-off between balance of stability and plasticity and energy efficiency. Introducing mechanisms to reduce the magnitude of metaplasticity coefficients based on neural activity can also potentially improve the balance of stability and plasticity \cite{nick_thesis}.  

\section*{Discussion}
Emerging nonvolatile memristor devices offer immense opportunities for energy-efficient continual learning on the edge. However, their limited precision and high variability may constrain their suitability for continual learning solutions. 
Our research takes inspiration from the computational principles of the brain to devise a continual learning mechanism suited to non-ideal low-precision memory devices. Computations in the brain are characterized by stochastic and low-precision synapses \cite{syn_low_precision1, syn_low_precision2}. Research in computational memory models demonstrates that stochastic binary synapses with low-precision metaplastic states can help memory retention \cite{fusi2005cascade}. Our work draws a parallel between the low-precision hidden metaplastic states of a cascade metaplastic synapse and the conductance levels of a memristor device. Abstracting this concept, we interpret the plasticity of weights as their probability of update rather than the magnitude of the update. This interpretation distinguishes the proposed work from previous continual learning models with metaplasticity. Previous work has also explored probabilistic updates in memristor weights considering learning settings with independent and identically distributed (i.i.d.) data \cite{misba_domain, aicas_stochastic}. In a continual learning setting, where the distribution of data changes, such mechanisms are prone to catastrophic forgetting. In contrast, probabilistic metaplasticity effectively alleviates catastrophic forgetting by dynamically adjusting the update probability of weights based on their importance.

In this work, we demonstrated the efficacy of probabilistic metaplasticity in a spiking network trained with error threshold-based eRBP, where weights were realized with a Hafnium oxide-based memristor. However, the proposed mechanism is not limited to the setting above and can be explored in the context of different network architectures and memory devices. The proposed work naturally extends to other surrogate gradient-based training algorithms for spiking networks, where the local error at the neurons can be estimated \cite{decolle}. Additionally, it can be applied to deep neural networks with low-precision weights updated using stochastic rounding \cite{stochastic_rounding}.
Incorporating probabilistic metaplasticity with gradient accumulation-based training is also possible. In this setting, probabilistic metaplasticity will eliminate the need to multiply metaplasticity factors with gradients. However, the high frequency of memory access may lead to a lower gain in energy consumption.
The proposed mechanism also extends to networks that realize low-precision weights with other memory devices. It shows promise to enable continual learning in neuromorphic accelerators with low-precision weights \cite{Loihi}.
Recent research has explored probabilistic training algorithms for emerging low-precision memory devices in deep neural networks considering learning settings with independent and identically distributed data \cite{misba_domain, aicas_stochastic}. 
Probabilistic metaplasticity can be explored in these networks to alleviate catastrophic forgetting during sequential learning with non-stationary data distribution.

We explored the effect of different weight resolutions by employing multi-memristor weights alongside single-memristor weights. Although we observed improved performance with higher resolution weights, it increases the area footprint of memristor crossbars. Improving the resolution of a single memristor device can lead to better performance with lower area overhead. Recent research has shown memristor devices with up to $\sim$ 6-bit precision \cite{mem_64_level}. In addition to the resolution of conductance modulation, the nonlinearity and variability in memristor devices can affect the performance of error threshold-based training. Due to these non-ideal characteristics, weight updates vary in response to the same local error depending on the current conductance level of the memristor. 
To minimize this effect, we restricted weight mapping to only ten low-resistance levels of the 1T1R device. This choice avoids the higher variability observed in the high-resistance levels along with the substantial difference in conductance between the highest low-resistance and high-resistance levels during weight updates\cite{liehr2020impact}. However, it should be noted that choosing low-resistance levels increases parasitic voltage drop and sneak currents, which can also introduce nonlinearity and degrade the network performance. Similarly, extreme non-uniformity in the distribution of conductance levels can potentially degrade the performance of error threshold-based training. 
Future studies incorporating optimized circuits and system demonstrations of the proposed mechanism can further investigate the effect of these non-idealities.

This work limits the exploration of probabilistic metaplasticity in fully-connected spiking networks. Future research can expand the proposed mechanism to larger networks and tackle more complex continual learning tasks with longer sequences. It would be interesting to extend the proposed mechanism to convolutional spiking networks where weights are shared among inputs. We identified a trade-off between memory overhead and the network's ability to balance stability and plasticity when metaplasticity coefficients are shared among network weights. Further theoretical analysis could offer valuable insights and pave the way for more efficient continual learning solutions.
Further investigations into the contrasting learning dynamics when training the network based on error accumulated at neurons instead of gradients can also lead to a deeper understanding of the underlying mechanisms. Our empirical observations on continual learning tasks did not show any significant differences in performance (Supplementary Table S6). However, we note that gradient accumulation may allow for more gradual weight changes over time as the gradients can be accumulated over samples. During error threshold-based training,  integrating error across samples may lead to incorrect attribution, especially when weights are updated based on immediate input activity. For settings that require gradual updates, tracing input activity over time or using low-precision counters to track smaller changes may prove beneficial. Error-triggered training is promising for resource-constrained platforms as it reduces memory overhead and merits robust study under various conditions.

In summary, we propose probabilistic metaplasticity for continual learning with low-precision memristor weights. The proposed approach consolidates important weights by modulating the probability of weight update rather than its magnitude. This eliminates the need for auxiliary high-precision memory for weight consolidation, leading to energy-efficient training with fewer memory accesses and computations. Benchmark evaluations demonstrated that probabilistic metaplasticity effectively prevents catastrophic forgetting in the online continual learning scenario with no task supervision, satisfying the criteria for autonomous applications. Moreover, it enables a spiking network with low-precision memristor weights to achieve accuracies equivalent to state-of-the-art continual learning models with high-precision weights while requiring lower memory overhead. Probabilistic metaplasticity can thus lead to efficient solutions for autonomous continual learning on edge.

\section*{Methods}
\subsection*{Memristor device}
We evaluate the proposed probabilistic metaplasticity model considering Hafnium Oxide-based 1T1R cells as weights. The Resistive Random Access Memory (RRAM) device consists of a $\text{TiN/Ti/HfO}_\text{2}/\text{Ti}$ stack connected to a transistor in the IBM 65nm 10LPe CMOS/Memristor process \cite{liehr2020impact}. The device can be used in binary mode by setting and resetting it to high and low conductive states. During SET operation, modulating the compliance currents by controlling the gate voltage of the transistor gives rise to 10 distinct levels in the high-conductive state, with mean conductance in the range of $\sim$ 40 $\mu$S - 283 $\mu$S with an average conductance resolution of $\sim$ 27 $\mu\text{S}$ \cite{liehr2020impact}. When used as a synaptic weight, the resolution of this device ( calculated as (maximum conductance - minimum conductance)/ average conductance resolution) is close to 3 bits of precision. In the multi-memristor weight, due to the parallel combination of memristors, the maximum and minimum conductance of the weight increases. However, since only one device is updated during training, the average conductance resolution during weight update is the same as a single memristor weight. We use scaling to map the multi-memristor weights to the desired range. With four and seven memristors per weight, the weight precision is close to 4 and 6 bits. 

\subsection*{Spiking neural network training algorithm}
We train the multilayer spiking neural network with event-driven random backpropagation (eRBP) combined with error threshold, which uses surrogate gradients \cite{neftci2017event}. This section describes the eRBP training rule. The input stimuli to the network are encoded as Poisson spike trains, where the spike rate is proportional to the strength of the input. The spike train propagates through the weights and accumulates as the input current to the Leaky Integrate and Fire neurons. For the $j^{\text{th}}$ neuron, the accumulated current can be expressed as - 

\begin{equation}
    I_j(t + 1) = I_j(t) + \frac{{\Delta}t}{\tau_{\text{syn}}} \Big(\sum_{i=1}^N  {w}_{ij} S_i(t) - I_j(t)\Big).
    \label{eq:current}
\end{equation}

Here $S_{i}(t)$ is the input spike at the $i^{\text{th}}$ pre-synaptic neuron at timestep $t$, $N$ is the total number of input neurons, $\tau_{\text{syn}}$ is the synaptic time constant, and $w_{ij}$ is the synaptic weight between $i^{\text{th}}$  and $j^{\text{th}}$ pre- and post-synaptic neurons. This current is integrated at the neuron membrane potential as shown in Equation \ref{eq:neuron}, where  $\tau_{\text{mem}}, R, C$ and $V_{\text{rest}}$ are the membrane time constant, membrane resistance, capacitance and the resting potential of the neuron.
 
\begin{equation}
    V_j(t+1) = V_j(t) + \frac{{\Delta}t}{\tau_{\text{mem}}}\bigg(\Big(V_{\text{rest}}-V_j(t)\Big)+I_j(t)R\bigg)
    \label{eq:neuron}
\end{equation}

When the membrane potential reaches its threshold $V_{\text{th}}$, the neuron emits a spike $S_j$, and its membrane potential is reset to the resting potential. A refractory period follows every time a neuron spikes, during which the neuron remains inactive. 
 
The spike train progresses through the network to produce the output layer spike train $S^{\text{out}}$. It is compared to the target spike train L to produce the error current $I^{\text{err}}$ -
\begin{equation}
    {I}^{\text{err}}(t) = {S}^{\text{out}}(t) - {L}(t).
    \label{eq:Ierr}
\end{equation}

Two sets of error neurons, for false positive and false negative errors, spike in response to $I^{\text{err}}$. The output spikes of error neurons ${S}^{\text{fp/fn}}$  are used to compute the error in the output and hidden layers. At the output layer $o$, the error at the $j^{\text{th}}$ neuron is -

\begin{equation}
    E^{o}_j(t) = {S}^{\text{fp}}_j(t) - {S}^{\text{fn}}_j(t).
\end{equation}

While for the hidden layer $h$, the error is backpropagated to the $i^{\text{th}}$ neuron by -

\begin{equation}
    E^{h}_i(t) = \sum_{j=1}^{n_{\text{out}}} \Big({w}^{\text{fp},h}_{ij}{S}^{\text{fp}}_j(t) - {w}^{\text{fn},h}_{ij}{S}^{\text{fn}}_j(t)\Big),
    \label{eq:errfb}
\end{equation}

\noindent where $n_{\text{out}}$ is the number of output neurons, and ${w}^{\text{fp},h}_{ij}$ and ${w}^{\text{fn},h}_{ij}$ are random feedback weights from error neurons encoding false positives and false negatives, respectively.

The neurons in the hidden and output layers integrate the errors as -
\begin{equation}
    U(t+1) = U(t) + \frac{{\Delta}t}{\tau_{\text{mem}}}\Big(-U(t) + E(t)R\Big).
    \label{eq:Ublock}
\end{equation}

As the dendritic compartment accumulates the error, the weights are updated in proportion to it whenever there is an active pre-synaptic spike, and the post-synaptic current is within a range. The update in weight can be expressed as -

\begin{equation}
    \Delta w_{ij} =  -\eta S_iU_j\Theta(I_j),
\end{equation}

where $\eta$ is the learning rate, $S_i$ is the pre-synaptic spike, $U_j$ is the error integrated at the dendritic compartment of the $j^{\text{th}}$ post-synaptic neuron, and $\Theta$ is a boxcar function, which indicates that the current magnitude at post-synaptic neuron $j$ falls within the threshold. 

\subsection*{Activity-dependent metaplasticity with gradient accumulation for continual learning}

Metaplasticity in neural networks is usually abstracted with a metaplastic factor, which modulates the magnitude of weight gradients \cite{laborieux2021synaptic, TACOS1}. The factor assumes a low value for important weight, which restricts subsequent changes and preserves previously learned information. In this work, we adopt activity-dependent metaplasticity \cite{TACOS1} for continual learning with gradient accumulation. According to this mechanism, the importance of a weight is determined by its magnitude and the activity of the adjacent neurons. The function is expressed as - 

\begin{equation}
    f(m_{ij},w_{ij}) = e^{-|m_{ij}w_{{ij}}|}
    \label{eq:meta}
\end{equation}

Here, $w_{ij}$ is the weight value, and $m_{ij}$ is a metaplasticity coefficient assigned to a weight that captures the activity of its adjacent neurons. It is updated by computing the trace of the neuron activity as -

\begin{equation}
    m_{ij}(t+1) =
\begin{cases}
    m_{ij}(t) + {\Delta}m, & \textrm{~if} \: X^{\text{tr}}_{j} \ge m^{\text{pre}}_{\text{th}} \& X^{\text{tr}}_{i} \ge m^{\text{post}}_{\text{th}} \\
    m_{ij}(t), & \textrm{~otherwise}
\end{cases},
\end{equation} 

The weight gradients are modulated as -

\begin{equation}
    \Delta w_{ij} =  -\eta S_iU_j\Theta(I_j)f(m_{ij},w_{ij})
\end{equation}

The required weight updates are highly precise and cannot be directly incorporated into low-precision memristor devices. Therefore, the weight gradients are accumulated in high-precision memory during sample spike-train presentations. Whenever the gradient reaches a threshold corresponding to the memristor resolution, the weights are updated. The gradients are reset after each sample presentation. The complete training procedure is described in Supplementary Algorithm S1.

\nolinenumbers
\section*{Data availability statement}
All datasets used in this work (MNIST \cite{MNIST_dataset}, Fashion-MNIST \cite{xiao2017fashion} and CIFAR-10 \cite{cifar10}) are publicly available.

\section*{Code availability statement}
The source code supporting the figures and experiments in this work is available from the corresponding author upon reasonable request.

\section*{Acknowledgements}
This effort is partially supported by NSF EFRI BRAID Award \#2317706,  NSF NAIAD Award \#2332744 and Air Force Research Laboratory under agreement number FA8750-20-2-1003 through BAA FA8750-19-S-7010. We thank Dr. Maximilian Liehr and Dr. Nathaniel Cady for helpful discussions on the RRAM device characteristics. We thank Dr. Abdullah Zyarah for helpful discussions and feedback. The views and conclusions contained herein are those of the authors and should not be interpreted as necessarily representing the official policies or endorsements, either expressed or implied, of NSF, Air Force Research Laboratory or the U.S. Government.

\section*{Author contributions statement}
F.Z. conceived the idea, conducted the experiments, and wrote the manuscript. V.K. contributed to the energy estimation for the digital components and memory models. N.S. helped develop the experiments. D.K. oversaw the development and execution of the research and offered critical feedback. All authors reviewed the manuscript.

\section*{Additional information}

\section*{Competing interests}
The authors declare no competing interests.

\section*{Figure legends}
\begin{itemize}
    \item \textbf{Figure 1: Probabilistic metaplasticity with error threshold-based training.} \textbf{a.} Spiking network trained on error threshold where the network weights are realized with 1T1R memristor crossbar array. The dendritic compartments in the hidden and output layer neurons integrate the error. When the error reaches the threshold $U_{\text{th}}$, the memristor weights are updated to the next higher (negative error) or lower (positive error) conductance level with probability $p_{\text{update}}$. \textbf{b.} Mean and standard deviation of resistance levels versus programming compliance current for the 1T1R memristor device (shown in inset) adopted in this work \cite{liehr2020impact}. \textbf{c.} Update probability of weights for different values of the metaplasticity coefficient $m$ and the weights $w$. $m$ is positively associated with the activity level of adjacent neurons. We see that weights with highly active adjacent neurons (high $m$) and weight magnitude (high $|w|$) lead to low update probability.
    \item \textbf{Figure 2: Evolution of task accuracies with sequential training on the split-MNIST benchmark.} The x-axis shows the latest task the network has learned. We evaluate the performance of task $n$ only after the network encounters it and denote the accuracies as 0 before that. \textbf{a.} With no probabilistic metaplasticity, the network learns the current task well, but forgets the initial tasks after sequentially learning multiple tasks. \textbf{b.} Probabilistic metaplasticity with low activity threshold leads to high rigidity in the network, so it remembers previous tasks but cannot learn the last task. \textbf{c.} High activity threshold can lead to loss in previous task accuracy while the network remains plastic to learn the new tasks. \textbf{d.} Optimized activity threshold balances plasticity and rigidity such that the network maintains high initial task accuracies while maintaining the ability to learn new tasks.
    \item \textbf{Figure 3: Effect of number of training samples on continual learning.} The mean accuracy across tasks shows degradation as the average number of samples per task reduces.
    \item \textbf{Figure 4: Mixed-signal architecture and computational flow for the continual learning mechanisms}. We consider a mixed-signal architecture where memristor crossbars carry out the forward pass of the spiking model and the error computation, arithmetic and logical operations are carried out in the digital domain. The figure shows the computational flow for continual learning with probabilistic metaplasticity (left) and activity-dependent metaplasticity with gradient accumulation (right).
    \item \textbf{Energy consumption and stability-plasticity tradeoff.} \textbf{a.} Energy consumption per sample and its breakdown during the parameter update phase for activity-dependent metaplasticity with gradient accumulation and probabilistic metaplasticity with individual and shared metaplasicity coefficients.  \textbf{b.} Distribution of the product of metaplasticity coefficient $m$ and weight magnitude $|w|$ in the hidden layer with probabilistic metaplasticity. We see that as more weights share $m$, lower fraction of weights assume low $|mw|$ value. This indicates reduced plasticity in the network. \textbf{c.} Mean accuracy across tasks vs. the energy consumption per sample for the split-MNIST task. We observe a tradeoff between continual learning performance and energy consumption as the metaplasticity coefficients are shared across weights.
\end{itemize}

\begin{table}[h]

\caption{The mean accuracies across tasks with standard deviations of different continual learning models on the split-MNIST and split-Fashion MNIST tasks after sequential training in the domain-IL scenario.}\label{tab:cont_acc}%
 \adjustbox{max width=\textwidth}{
\begin{tabular}{@{}llll@{}}
\toprule
Model & Split-MNIST  & Split-Fashion MNIST & Normalized Memory  \\
&   &  &  Overhead \\
\midrule
Baseline  & 60.69 \% $\pm$ 0.6 & 75.52\% $\pm$ 1.31  & 0 kB\\ 
LwF \cite{li2017learning}  & 71.50\% $\pm$ 1.63 & 71.02\% $\pm$ 0.46 & 628.8 kB\\ 

MAS \cite{Aljundi_MAS}  & 66.42\% $\pm$ 2.47 & 68.57\% $\pm$ 6.85 & $\sim$ 1.2 MB\\  

BGD \cite{zeno2018task}  & 80.44\% $\pm$ 0.45 & 89.73\% $\pm$ 0.88 & $\sim$ 1.2 MB \\ 

SS \cite{SS}  & \textbf{82.90\% $\pm$ 0.01} & 91.98\% $\pm$ 0.12 & 628.8 kB\\ 

TACOS \cite{TACOS1}  & 82.56\% $\pm$ 1.12 & \textbf{93.22\% $\pm$ 0.22} & $\sim$ 0.9 MB\\
Probabilistic Metaplasticity  & &  & \\
 $n_{\text{mem}} =1$  & 81.14\% $\pm$ 1.54 & 92.56\% $\pm$ 1.49 & $\sim$ 0.3 MB\\
 $n_{\text{mem}} =2$  & 82.67\% $\pm$ 1.21 & 92.42\% $\pm$ 0.63 & $\sim$ 0.3 MB\\
$n_{\text{mem}} = 7$ & \textbf{83.69\% $\pm$ 0.78} & \textbf{93.23\% $\pm$ 0.14} & $\sim$ 0.3 MB\\
Activity-dependent Metaplasticity
 &  &  & \\

w/ Gradient Accumulation ($n_{\text{mem}} = 7$) & 83.18\% $\pm$ 0.42 & 92.33\% $\pm$ 0.27 & $\sim$ 0.9 MB\\
\bottomrule
\end{tabular}}
\label{tab:}
\end{table}

\begin{table}[h]

    \caption{Performance of a spiking network on the split-MNIST task in different settings. The table shows the mean and standard deviation of the individual task accuracies and the mean accuracy across tasks over 5 runs after sequential training.}

    \begin{tabular}{lcccccc}
        \toprule
         & Baseline & Random  & \multicolumn{3}{c}{Decaying Probabilistic } & Probabilistic \\
         & &  Consolidation & \multicolumn{3}{c}{ Plasticity} &  Metaplasticity \\
        \cmidrule(lr){4-6}
        & &  & (Factor=2) & (Factor=5) & (Factor=10) & \\
        \midrule
        Task 1 & 44.31 $\pm$ 3.80 & 48.01 $\pm$ 2.77 & 47.05 $\pm$ 1.63 & 50.11 $\pm$ 1.89 & 73.58 $\pm$ 2.80 & 84.95 $\pm$ 3.42 \\
        Task 2 & 53.67 $\pm$ 1.42 & 57.14 $\pm$ 0.61 & 56.39 $\pm$ 0.62 & 94.31 $\pm$ 0.27 & 94.76 $\pm$ 0.43 & 90.20 $\pm$ 1.96 \\
        Task 3 & 8.63 $\pm$ 1.04 & 17.73 $\pm$ 1.09 & 22.42 $\pm$ 1.53 & 90.51 $\pm$ 0.40 & 84.59 $\pm$ 0.59 & 72.06 $\pm$ 2.60 \\
        Task 4 & 88.31 $\pm$ 1.24 & 96.24 $\pm$ 0.55 & 96.68 $\pm$ 0.44 & 93.59 $\pm$ 0.91 & 82.72 $\pm$ 1.39 & 94.73 $\pm$ 0.94 \\
        Task 5 & 97.52 $\pm$ 0.32 & 95.73 $\pm$ 0.62 & 95.17 $\pm$ 0.38 & 48.85 $\pm$ 1.66 & 40.44 $\pm$ 1.16 & 76.54 $\pm$ 2.60 \\
        \midrule
        Mean & 58.49 $\pm$ 0.73 & 62.97 $\pm$ 0.56 & 63.54 $\pm$ 0.36 & 75.47 $\pm$ 0.32 & 75.22 $\pm$ 0.79 & 83.70 $\pm$ 0.78 \\
        \bottomrule
    \end{tabular}%
    
    \label{tab:control_exp}

\end{table}

\begin{table}[h]
    \centering
    \caption{Evaluation of a spiking neural network with probabilistic metaplasticity and TACOS on the split-CIFAR-10 task in the domain-IL scenario. Probabilistic metaplasticity is evaluated considering multi-memristor weights $n_{mem}$ = 7 and TACOS considers full-precision weights. The table lists the mean and standard deviation of the individual task accuracies and the mean accuracy across tasks over 5 runs after sequential training.}
    \begin{tabular}{lccc}
        \toprule
        Task & Baseline & Probabilistic Metaplasticity & TACOS\cite{TACOS1}\\
        \midrule
         
        Class 0,1  & 86.64 $\pm$ 1.86 & 93.52 $\pm$ 1.36 & 90.84 $\pm$ 0.36\\
        Class 2,3 & 59.81 $\pm$ 3.69 & 73.11 $\pm$ 4.80 & 71.27  $\pm$ 1.08\\
        Class 4,5  & 63.93 $\pm$ 3.08 & 80.87 $\pm$ 4.07 & 80.01  $\pm$0.93\\
        Class 6,7  & 82.03 $\pm$ 5.75 & 84.90 $\pm$ 1.61 & 88.83 $\pm$ 0.57\\
        Class 8,9  & 96.96 $\pm$ 0.80 & 94.71 $\pm$ 0.21 & 97.64  $\pm$ 0.23\\
        \bottomrule
        Mean  & 77.87 $\pm$ 2.77 & 85.42 $\pm$ 2.11  & 85.72  $\pm$ 0.55\\

        \bottomrule
    \end{tabular}
\label{tab:CIFAR-10}
\end{table}

\begin{table}[h]
    \centering
    \caption{Evaluation of probabilistic metaplasticity and neural activity-dependent weight consolidation models on the split-MNIST and split-Fashion MNIST tasks in the domain-IL scenario.}
    
    \begin{tabular}{lccc}
        \toprule
        Model & Split-MNIST  & Split-FMNIST & Normalized Memory\\
        &   &  &  Overhead\\
        \midrule
        Probabilistic Metaplasticity & & &\\
        Individual m  & 83.70 $\pm$ 0.78 & 93.23 $\pm$ 0.14 & $\sim$ 0.3 MB\\
        
        Neuron-shared m  & 82.07 $\pm$ 0.84 & 92.72 $\pm$ 0.30 & $\sim$ 0.4 kB\\
        
        Layer-shared m  & 79.74 $\pm$ 1.21 & 92.01 $\pm$ 0.49  &  4 Bytes\\
        
        NAI \cite{NAI} & 68.35 $\pm$ 1.34 & 68.82 $\pm$ 1.15 & $\sim$ 0.8 kB \\
        NEO \cite{NEO} & 78.14 $\pm$ 2.23 & 86.82 $\pm$ 0.60 & $\sim$ 0.3 MB\\

        \bottomrule
    \end{tabular}
\label{tab:neu_table}
\end{table}


\begin{thebibliography}{10}
\urlstyle{rm}
\expandafter\ifx\csname url\endcsname\relax
  \def\url#1{\texttt{#1}}\fi
\expandafter\ifx\csname urlprefix\endcsname\relax\def\urlprefix{URL }\fi
\expandafter\ifx\csname doiprefix\endcsname\relax\def\doiprefix{DOI: }\fi
\providecommand{\bibinfo}[2]{#2}
\providecommand{\eprint}[2][]{\url{#2}}

\bibitem{hayes2022online}
\bibinfo{author}{Hayes, T.~L.} \& \bibinfo{author}{Kanan, C.}
\newblock \bibinfo{journal}{\bibinfo{title}{{Online Continual Learning for Embedded Devices}}}.
\newblock {\emph{\JournalTitle{arXiv preprint arXiv:2203.10681}}}  (\bibinfo{year}{2022}).

\bibitem{8778327}
\bibinfo{author}{Kukreja, N.} \emph{et~al.}
\newblock \bibinfo{title}{{Training on the Edge: The why and the how}}.
\newblock In \emph{\bibinfo{booktitle}{2019 IEEE International Parallel and Distributed Processing Symposium Workshops (IPDPSW)}}, \bibinfo{pages}{899--903}, \doiprefix\url{10.1109/IPDPSW.2019.00148} (\bibinfo{year}{2019}).

\bibitem{dally2022model}
\bibinfo{author}{Dally, W.}
\newblock \bibinfo{journal}{\bibinfo{title}{{On the Model of Computation: Point}}}.
\newblock {\emph{\JournalTitle{Communications of the ACM}}} \textbf{\bibinfo{volume}{65}}, \bibinfo{pages}{30--32} (\bibinfo{year}{2022}).

\bibitem{FRENCH1999128}
\bibinfo{author}{French, R.~M.}
\newblock \bibinfo{journal}{\bibinfo{title}{Catastrophic forgetting in connectionist networks}}.
\newblock {\emph{\JournalTitle{Trends in Cognitive Sciences}}} \textbf{\bibinfo{volume}{3}}, \bibinfo{pages}{128--135}, \doiprefix\url{https://doi.org/10.1016/S1364-6613(99)01294-2} (\bibinfo{year}{1999}).

\bibitem{verwimp2023continual}
\bibinfo{author}{Verwimp, E.} \emph{et~al.}
\newblock \bibinfo{journal}{\bibinfo{title}{{Continual Learning: Applications and the Road Forward}}}.
\newblock {\emph{\JournalTitle{arXiv preprint arXiv:2311.11908}}}  (\bibinfo{year}{2023}).

\bibitem{kudithipudi2023design}
\bibinfo{author}{Kudithipudi, D.} \emph{et~al.}
\newblock \bibinfo{journal}{\bibinfo{title}{{Design principles for lifelong learning AI accelerators}}}.
\newblock {\emph{\JournalTitle{Nature Electronics}}} \bibinfo{pages}{1--16} (\bibinfo{year}{2023}).

\bibitem{7155485}
\bibinfo{author}{Balasubramanian, V.}
\newblock \bibinfo{journal}{\bibinfo{title}{{Heterogeneity and Efficiency in the Brain}}}.
\newblock {\emph{\JournalTitle{Proceedings of the IEEE}}} \textbf{\bibinfo{volume}{103}}, \bibinfo{pages}{1346--1358}, \doiprefix\url{10.1109/JPROC.2015.2447016} (\bibinfo{year}{2015}).

\bibitem{biological_underpinnings}
\bibinfo{author}{Kudithipudi, D.} \emph{et~al.}
\newblock \bibinfo{journal}{\bibinfo{title}{Biological underpinnings for lifelong learning machines}}.
\newblock {\emph{\JournalTitle{Nature Machine Intelligence}}} \textbf{\bibinfo{volume}{4}}, \bibinfo{pages}{196--210}, \doiprefix\url{10.1038/s42256-022-00452-0} (\bibinfo{year}{2022}).

\bibitem{abraham_meta}
\bibinfo{author}{Abraham, W.~C.} \& \bibinfo{author}{Bear, M.~F.}
\newblock \bibinfo{journal}{\bibinfo{title}{Metaplasticity: the plasticity of synaptic plasticity}}.
\newblock {\emph{\JournalTitle{Trends in Neurosciences}}} \textbf{\bibinfo{volume}{19}}, \bibinfo{pages}{126--130}, \doiprefix\url{https://doi.org/10.1016/S0166-2236(96)80018-X} (\bibinfo{year}{1996}).

\bibitem{Benna2016}
\bibinfo{author}{Benna, M.~K.} \& \bibinfo{author}{Fusi, S.}
\newblock \bibinfo{journal}{\bibinfo{title}{Computational principles of synaptic memory consolidation}}.
\newblock {\emph{\JournalTitle{Nature Neuroscience}}} \textbf{\bibinfo{volume}{19}}, \bibinfo{pages}{1697--1706}, \doiprefix\url{10.1038/nn.4401} (\bibinfo{year}{2016}).

\bibitem{TACOS1}
\bibinfo{author}{Soures, N.}, \bibinfo{author}{Helfer, P.}, \bibinfo{author}{Daram, A.}, \bibinfo{author}{Pandit, T.} \& \bibinfo{author}{Kudithipudi, D.}
\newblock \bibinfo{title}{{TACOS: Task Agnostic Continual Learning in Spiking Neural Network}}.
\newblock In \emph{\bibinfo{booktitle}{Theory and Foundation of Continual Learning Workshop at ICML’2021}} (\bibinfo{year}{July 2021}).

\bibitem{ewc}
\bibinfo{author}{Kirkpatrick, J.} \emph{et~al.}
\newblock \bibinfo{journal}{\bibinfo{title}{Overcoming catastrophic forgetting in neural networks}}.
\newblock {\emph{\JournalTitle{Proceedings of the National Academy of Sciences}}} \textbf{\bibinfo{volume}{114}}, \bibinfo{pages}{3521--3526}, \doiprefix\url{10.1073/pnas.1611835114} (\bibinfo{year}{2017}).
\newblock \eprint{https://www.pnas.org/doi/pdf/10.1073/pnas.1611835114}.

\bibitem{SI}
\bibinfo{author}{Zenke, F.}, \bibinfo{author}{Poole, B.} \& \bibinfo{author}{Ganguli, S.}
\newblock \bibinfo{title}{{Continual Learning Through Synaptic Intelligence}}.
\newblock In \bibinfo{editor}{Precup, D.} \& \bibinfo{editor}{Teh, Y.~W.} (eds.) \emph{\bibinfo{booktitle}{Proceedings of the 34th International Conference on Machine Learning}}, vol.~\bibinfo{volume}{70} of \emph{\bibinfo{series}{Proceedings of Machine Learning Research}}, \bibinfo{pages}{3987--3995} (\bibinfo{publisher}{PMLR}, \bibinfo{year}{2017}).

\bibitem{pmlr-v80-kaplanis18a}
\bibinfo{author}{Kaplanis, C.}, \bibinfo{author}{Shanahan, M.} \& \bibinfo{author}{Clopath, C.}
\newblock \bibinfo{title}{{Continual Reinforcement Learning with Complex Synapses}}.
\newblock In \bibinfo{editor}{Dy, J.} \& \bibinfo{editor}{Krause, A.} (eds.) \emph{\bibinfo{booktitle}{Proceedings of the 35th International Conference on Machine Learning}}, vol.~\bibinfo{volume}{80} of \emph{\bibinfo{series}{Proceedings of Machine Learning Research}}, \bibinfo{pages}{2497--2506} (\bibinfo{publisher}{PMLR}, \bibinfo{year}{2018}).

\bibitem{laborieux2021synaptic}
\bibinfo{author}{Laborieux, A.}, \bibinfo{author}{Ernoult, M.}, \bibinfo{author}{Hirtzlin, T.} \& \bibinfo{author}{Querlioz, D.}
\newblock \bibinfo{journal}{\bibinfo{title}{Synaptic metaplasticity in binarized neural networks}}.
\newblock {\emph{\JournalTitle{Nature Communications}}} \textbf{\bibinfo{volume}{12}}, \bibinfo{pages}{2549} (\bibinfo{year}{2021}).

\bibitem{SNN_benefits}
\bibinfo{author}{Han, B.}, \bibinfo{author}{Sengupta, A.} \& \bibinfo{author}{Roy, K.}
\newblock \bibinfo{title}{{On the Energy Benefits of Spiking Deep Neural Networks: A Case Study}}.
\newblock In \emph{\bibinfo{booktitle}{2016 International Joint Conference on Neural Networks (IJCNN)}}, \bibinfo{pages}{971--976}, \doiprefix\url{10.1109/IJCNN.2016.7727303} (\bibinfo{year}{2016}).

\bibitem{brain_inspired_memristor}
\bibinfo{author}{Yu, S.}
\newblock \bibinfo{journal}{\bibinfo{title}{{Neuro-Inspired Computing with Emerging Nonvolatile memory}}}.
\newblock {\emph{\JournalTitle{Proceedings of the IEEE}}} \textbf{\bibinfo{volume}{106}}, \bibinfo{pages}{260--285}, \doiprefix\url{10.1109/JPROC.2018.2790840} (\bibinfo{year}{2018}).

\bibitem{Xia2019}
\bibinfo{author}{Xia, Q.} \& \bibinfo{author}{Yang, J.~J.}
\newblock \bibinfo{journal}{\bibinfo{title}{Memristive crossbar arrays for brain-inspired computing}}.
\newblock {\emph{\JournalTitle{Nature Materials}}} \textbf{\bibinfo{volume}{18}}, \bibinfo{pages}{309--323}, \doiprefix\url{10.1038/s41563-019-0291-x} (\bibinfo{year}{2019}).

\bibitem{time}
\bibinfo{author}{Cheng, M.} \emph{et~al.}
\newblock \bibinfo{title}{{TIME: A Training-in-Memory Architecture for Memristor-Based Deep Neural Networks}}.
\newblock In \emph{\bibinfo{booktitle}{Proceedings of the 54th Annual Design Automation Conference 2017}}, DAC '17, \doiprefix\url{10.1145/3061639.3062326} (\bibinfo{publisher}{Association for Computing Machinery}, \bibinfo{address}{New York, NY, USA}, \bibinfo{year}{2017}).

\bibitem{liehr2020impact}
\bibinfo{author}{Liehr, M.}, \bibinfo{author}{Hazra, J.}, \bibinfo{author}{Beckmann, K.}, \bibinfo{author}{Rafiq, S.} \& \bibinfo{author}{Cady, N.}
\newblock \bibinfo{title}{{Impact of Switching Variability of 65nm CMOS Integrated Hafnium Dioxide-based ReRAM Devices on Distinct Level Operations}}.
\newblock In \emph{\bibinfo{booktitle}{2020 IEEE International Integrated Reliability Workshop (IIRW)}}, \bibinfo{pages}{1--4} (\bibinfo{organization}{IEEE}, \bibinfo{year}{2020}).

\bibitem{mem_64_level}
\bibinfo{author}{Park, J.} \emph{et~al.}
\newblock \bibinfo{journal}{\bibinfo{title}{{TiOx-Based RRAM Synapse With 64-Levels of Conductance and Symmetric Conductance Change by Adopting a Hybrid Pulse Scheme for Neuromorphic Computing}}}.
\newblock {\emph{\JournalTitle{IEEE Electron Device Letters}}} \textbf{\bibinfo{volume}{37}}, \bibinfo{pages}{1559--1562}, \doiprefix\url{10.1109/LED.2016.2622716} (\bibinfo{year}{2016}).

\bibitem{comp_memory}
\bibinfo{author}{Nandakumar, S.~R.} \emph{et~al.}
\newblock \bibinfo{journal}{\bibinfo{title}{{Mixed-Precision Deep Learning Based on Computational Memory}}}.
\newblock {\emph{\JournalTitle{Frontiers in Neuroscience}}} \textbf{\bibinfo{volume}{14}}, \doiprefix\url{10.3389/fnins.2020.00406} (\bibinfo{year}{2020}).

\bibitem{mp_cont_learning}
\bibinfo{author}{Li, Y.} \emph{et~al.}
\newblock \bibinfo{journal}{\bibinfo{title}{Mixed-precision continual learning based on computational resistance random access memory}}.
\newblock {\emph{\JournalTitle{Advanced Intelligent Systems}}} \textbf{\bibinfo{volume}{4}}, \bibinfo{pages}{2200026}, \doiprefix\url{https://doi.org/10.1002/aisy.202200026} (\bibinfo{year}{2022}).
\newblock \eprint{https://onlinelibrary.wiley.com/doi/pdf/10.1002/aisy.202200026}.

\bibitem{d2023synaptic}
\bibinfo{author}{D’Agostino, S.} \emph{et~al.}
\newblock \bibinfo{title}{{Synaptic metaplasticity with multi-level memristive devices}}.
\newblock In \emph{\bibinfo{booktitle}{2023 IEEE 5th International Conference on Artificial Intelligence Circuits and Systems (AICAS)}}, \bibinfo{pages}{1--5}, \doiprefix\url{10.1109/AICAS57966.2023.10168563} (\bibinfo{year}{2023}).

\bibitem{Lu2024}
\bibinfo{author}{Lu, A.} \emph{et~al.}
\newblock \bibinfo{journal}{\bibinfo{title}{{High-speed emerging memories for AI hardware accelerators}}}.
\newblock {\emph{\JournalTitle{Nature Reviews Electrical Engineering}}} \textbf{\bibinfo{volume}{1}}, \bibinfo{pages}{24--34}, \doiprefix\url{10.1038/s44287-023-00002-9} (\bibinfo{year}{2024}).

\bibitem{fusi2005cascade}
\bibinfo{author}{Fusi, S.}, \bibinfo{author}{Drew, P.~J.} \& \bibinfo{author}{Abbott, L.~F.}
\newblock \bibinfo{journal}{\bibinfo{title}{{Cascade Models of Synaptically Stored Memories}}}.
\newblock {\emph{\JournalTitle{Neuron}}} \textbf{\bibinfo{volume}{45}}, \bibinfo{pages}{599--611} (\bibinfo{year}{2005}).

\bibitem{metaplasticnet}
\bibinfo{author}{Zohora, F.~T.}, \bibinfo{author}{Karia, V.}, \bibinfo{author}{Daram, A.~R.}, \bibinfo{author}{Zyarah, A.~M.} \& \bibinfo{author}{Kudithipudi, D.}
\newblock \bibinfo{title}{{MetaplasticNet: Architecture with Probabilistic Metaplastic Synapses for Continual Learning}}.
\newblock In \emph{\bibinfo{booktitle}{2021 IEEE International Symposium on Circuits and Systems (ISCAS)}}, \bibinfo{pages}{1--5}, \doiprefix\url{10.1109/ISCAS51556.2021.9401262} (\bibinfo{year}{2021}).

\bibitem{meta_memory}
\bibinfo{author}{Crestani, A.~P.} \emph{et~al.}
\newblock \bibinfo{journal}{\bibinfo{title}{Metaplasticity contributes to memory formation in the hippocampus}}.
\newblock {\emph{\JournalTitle{Neuropsychopharmacology}}} \textbf{\bibinfo{volume}{44}}, \bibinfo{pages}{408--414}, \doiprefix\url{10.1038/s41386-018-0096-7} (\bibinfo{year}{2019}).

\bibitem{neftci2017event}
\bibinfo{author}{Neftci, E.~O.}, \bibinfo{author}{Augustine, C.}, \bibinfo{author}{Paul, S.} \& \bibinfo{author}{Detorakis, G.}
\newblock \bibinfo{journal}{\bibinfo{title}{{Event-Driven Random Back-Propagation: Enabling Neuromorphic Deep Learning Machines}}}.
\newblock {\emph{\JournalTitle{Frontiers in Neuroscience}}} \textbf{\bibinfo{volume}{11}}, \bibinfo{pages}{324} (\bibinfo{year}{2017}).

\bibitem{error-triggered}
\bibinfo{author}{Payvand, M.}, \bibinfo{author}{Fouda, M.~E.}, \bibinfo{author}{Kurdahi, F.}, \bibinfo{author}{Eltawil, A.~M.} \& \bibinfo{author}{Neftci, E.~O.}
\newblock \bibinfo{journal}{\bibinfo{title}{{On-Chip Error-Triggered Learning of Multi-Layer Memristive Spiking Neural Networks}}}.
\newblock {\emph{\JournalTitle{IEEE Journal on Emerging and Selected Topics in Circuits and Systems}}} \textbf{\bibinfo{volume}{10}}, \bibinfo{pages}{522--535}, \doiprefix\url{10.1109/JETCAS.2020.3040248} (\bibinfo{year}{2020}).

\bibitem{bias_col_paper}
\bibinfo{author}{Zyarah, A.~M.} \& \bibinfo{author}{Kudithipudi, D.}
\newblock \bibinfo{journal}{\bibinfo{title}{{Semi-Trained Memristive Crossbar Computing Engine with In Situ Learning Accelerator}}}.
\newblock {\emph{\JournalTitle{J. Emerg. Technol. Comput. Syst.}}} \textbf{\bibinfo{volume}{14}}, \doiprefix\url{10.1145/3233987} (\bibinfo{year}{2018}).

\bibitem{vandeVen2022}
\bibinfo{author}{van~de Ven, G.~M.}, \bibinfo{author}{Tuytelaars, T.} \& \bibinfo{author}{Tolias, A.~S.}
\newblock \bibinfo{journal}{\bibinfo{title}{Three types of incremental learning}}.
\newblock {\emph{\JournalTitle{Nature Machine Intelligence}}} \textbf{\bibinfo{volume}{4}}, \bibinfo{pages}{1185--1197}, \doiprefix\url{10.1038/s42256-022-00568-3} (\bibinfo{year}{2022}).

\bibitem{Hsu18_EvalCL}
\bibinfo{author}{Hsu, Y.-C.}, \bibinfo{author}{Liu, Y.-C.}, \bibinfo{author}{Ramasamy, A.} \& \bibinfo{author}{Kira, Z.}
\newblock \bibinfo{title}{{Re-evaluating Continual Learning Scenarios: A Categorization and Case for Strong Baselines}}.
\newblock In \emph{\bibinfo{booktitle}{NeurIPS Continual learning Workshop}} (\bibinfo{year}{2018}).

\bibitem{MNIST_dataset}
\bibinfo{author}{Lecun, Y.}, \bibinfo{author}{Bottou, L.}, \bibinfo{author}{Bengio, Y.} \& \bibinfo{author}{Haffner, P.}
\newblock \bibinfo{journal}{\bibinfo{title}{{Gradient-Based Learning Applied to Document Recognition}}}.
\newblock {\emph{\JournalTitle{Proceedings of the IEEE}}} \textbf{\bibinfo{volume}{86}}, \bibinfo{pages}{2278--2324}, \doiprefix\url{10.1109/5.726791} (\bibinfo{year}{1998}).

\bibitem{xiao2017fashion}
\bibinfo{author}{Xiao, H.}, \bibinfo{author}{Rasul, K.} \& \bibinfo{author}{Vollgraf, R.}
\newblock \bibinfo{journal}{\bibinfo{title}{{Fashion-MNIST: A Novel Image Dataset for Benchmarking Machine Learning Algorithms}}}.
\newblock {\emph{\JournalTitle{arXiv preprint arXiv:1708.07747}}}  (\bibinfo{year}{2017}).

\bibitem{Aljundi_MAS}
\bibinfo{author}{Aljundi, R.}, \bibinfo{author}{Babiloni, F.}, \bibinfo{author}{Elhoseiny, M.}, \bibinfo{author}{Rohrbach, M.} \& \bibinfo{author}{Tuytelaars, T.}
\newblock \bibinfo{title}{{Memory Aware Synapses: Learning what (not) to forget}}.
\newblock In \emph{\bibinfo{booktitle}{Proceedings of the European Conference on Computer Vision (ECCV)}} (\bibinfo{year}{2018}).

\bibitem{SS}
\bibinfo{author}{Schug, S.}, \bibinfo{author}{Benzing, F.} \& \bibinfo{author}{Steger, A.}
\newblock \bibinfo{journal}{\bibinfo{title}{Presynaptic stochasticity improves energy efficiency and helps alleviate the stability-plasticity dilemma}}.
\newblock {\emph{\JournalTitle{eLife}}} \textbf{\bibinfo{volume}{10}}, \bibinfo{pages}{e69884}, \doiprefix\url{10.7554/eLife.69884} (\bibinfo{year}{2021}).

\bibitem{zeno2018task}
\bibinfo{author}{Zeno, C.}, \bibinfo{author}{Golan, I.}, \bibinfo{author}{Hoffer, E.} \& \bibinfo{author}{Soudry, D.}
\newblock \bibinfo{journal}{\bibinfo{title}{{Task Agnostic Continual Learning Using Online Variational Bayes}}}.
\newblock {\emph{\JournalTitle{arXiv preprint arXiv:1803.10123}}}  (\bibinfo{year}{2018}).

\bibitem{li2017learning}
\bibinfo{author}{Li, Z.} \& \bibinfo{author}{Hoiem, D.}
\newblock \bibinfo{journal}{\bibinfo{title}{{Learning without Forgetting}}}.
\newblock {\emph{\JournalTitle{IEEE Transactions on Pattern Analysis and Machine Intelligence}}} \textbf{\bibinfo{volume}{40}}, \bibinfo{pages}{2935--2947}, \doiprefix\url{10.1109/TPAMI.2017.2773081} (\bibinfo{year}{2018}).

\bibitem{boybat2018neuromorphic}
\bibinfo{author}{Boybat, I.} \emph{et~al.}
\newblock \bibinfo{journal}{\bibinfo{title}{Neuromorphic computing with multi-memristive synapses}}.
\newblock {\emph{\JournalTitle{Nature Communications}}} \textbf{\bibinfo{volume}{9}}, \bibinfo{pages}{1--12} (\bibinfo{year}{2018}).

\bibitem{cifar10}
\bibinfo{author}{Krizhevsky, A.}
\newblock \bibinfo{journal}{\bibinfo{title}{Learning multiple layers of features from tiny images}}.
\newblock {\emph{\JournalTitle{Master's thesis, University of Tront}}}  (\bibinfo{year}{2009}).

\bibitem{NEO}
\bibinfo{author}{Daram, A.} \& \bibinfo{author}{Kudithipudi, D.}
\newblock \bibinfo{title}{{NEO: Neuron State Dependent Mechanisms for Efficient Continual Learning}}.
\newblock In \emph{\bibinfo{booktitle}{Proceedings of the 2023 Annual Neuro-Inspired Computational Elements Conference}}, NICE '23, \bibinfo{pages}{11–19}, \doiprefix\url{10.1145/3584954.3584960} (\bibinfo{publisher}{Association for Computing Machinery}, \bibinfo{address}{New York, NY, USA}, \bibinfo{year}{2023}).

\bibitem{NAI}
\bibinfo{author}{Kim, S.} \& \bibinfo{author}{Lee, S.}
\newblock \bibinfo{title}{{Continual Learning with Neuron Activation Importance}}.
\newblock In \bibinfo{editor}{Sclaroff, S.}, \bibinfo{editor}{Distante, C.}, \bibinfo{editor}{Leo, M.}, \bibinfo{editor}{Farinella, G.~M.} \& \bibinfo{editor}{Tombari, F.} (eds.) \emph{\bibinfo{booktitle}{Image Analysis and Processing -- ICIAP 2022}}, \bibinfo{pages}{310--321} (\bibinfo{publisher}{Springer International Publishing}, \bibinfo{address}{Cham}, \bibinfo{year}{2022}).

\bibitem{nick_thesis}
\bibinfo{author}{Soures, N.}
\newblock \emph{\bibinfo{title}{Lifelong Learning in Spiking Neural Networks Through Neural Plasticity}}.
\newblock Ph.D. thesis, \bibinfo{school}{Rochester Institute of Technology} (\bibinfo{year}{2023}).

\bibitem{syn_low_precision1}
\bibinfo{author}{Bartol, T.~M.} \emph{et~al.}
\newblock \bibinfo{journal}{\bibinfo{title}{{Hippocampal Spine Head Sizes Are Highly Precise}}}.
\newblock {\emph{\JournalTitle{bioRxiv}}} \doiprefix\url{10.1101/016329} (\bibinfo{year}{2015}).
\newblock \eprint{https://www.biorxiv.org/content/early/2015/03/11/016329.full.pdf}.

\bibitem{syn_low_precision2}
\bibinfo{author}{O'Connor, D.~H.}, \bibinfo{author}{Wittenberg, G.~M.} \& \bibinfo{author}{Wang, S. S.-H.}
\newblock \bibinfo{journal}{\bibinfo{title}{Graded bidirectional synaptic plasticity is composed of switch-like unitary events}}.
\newblock {\emph{\JournalTitle{Proceedings of the National Academy of Sciences}}} \textbf{\bibinfo{volume}{102}}, \bibinfo{pages}{9679--9684}, \doiprefix\url{10.1073/pnas.0502332102} (\bibinfo{year}{2005}).
\newblock \eprint{https://www.pnas.org/doi/pdf/10.1073/pnas.0502332102}.

\bibitem{misba_domain}
\bibinfo{author}{Misba, W.~A.}, \bibinfo{author}{Lozano, M.}, \bibinfo{author}{Querlioz, D.} \& \bibinfo{author}{Atulasimha, J.}
\newblock \bibinfo{journal}{\bibinfo{title}{{Energy Efficient Learning With Low Resolution Stochastic Domain Wall Synapse for Deep Neural Networks}}}.
\newblock {\emph{\JournalTitle{IEEE Access}}} \textbf{\bibinfo{volume}{10}}, \bibinfo{pages}{84946--84959}, \doiprefix\url{10.1109/ACCESS.2022.3196688} (\bibinfo{year}{2022}).

\bibitem{aicas_stochastic}
\bibinfo{author}{Zhang, Y.}, \bibinfo{author}{He, G.}, \bibinfo{author}{Tang, K.-T.} \& \bibinfo{author}{Wang, G.}
\newblock \bibinfo{title}{{On-chip Learning of Multilayer Perceptron Based on Memristors with Limited Multilevel States}}.
\newblock In \emph{\bibinfo{booktitle}{2019 IEEE International Conference on Artificial Intelligence Circuits and Systems (AICAS)}}, \bibinfo{pages}{11--12}, \doiprefix\url{10.1109/AICAS.2019.8771513} (\bibinfo{year}{2019}).

\bibitem{decolle}
\bibinfo{author}{Kaiser, J.}, \bibinfo{author}{Mostafa, H.} \& \bibinfo{author}{Neftci, E.}
\newblock \bibinfo{journal}{\bibinfo{title}{{Synaptic Plasticity Dynamics for Deep Continuous Local Learning (DECOLLE)}}}.
\newblock {\emph{\JournalTitle{Frontiers in Neuroscience}}} \textbf{\bibinfo{volume}{14}}, \doiprefix\url{10.3389/fnins.2020.00424} (\bibinfo{year}{2020}).

\bibitem{stochastic_rounding}
\bibinfo{author}{Gupta, S.}, \bibinfo{author}{Agrawal, A.}, \bibinfo{author}{Gopalakrishnan, K.} \& \bibinfo{author}{Narayanan, P.}
\newblock \bibinfo{title}{Deep learning with limited numerical precision}.
\newblock In \bibinfo{editor}{Bach, F.} \& \bibinfo{editor}{Blei, D.} (eds.) \emph{\bibinfo{booktitle}{Proceedings of the 32nd International Conference on Machine Learning}}, vol.~\bibinfo{volume}{37} of \emph{\bibinfo{series}{Proceedings of Machine Learning Research}}, \bibinfo{pages}{1737--1746} (\bibinfo{publisher}{PMLR}, \bibinfo{address}{Lille, France}, \bibinfo{year}{2015}).

\bibitem{Loihi}
\bibinfo{author}{Davies, M.} \emph{et~al.}
\newblock \bibinfo{journal}{\bibinfo{title}{{Loihi: A Neuromorphic Manycore Processor with On-Chip Learning}}}.
\newblock {\emph{\JournalTitle{IEEE Micro}}} \textbf{\bibinfo{volume}{38}}, \bibinfo{pages}{82--99}, \doiprefix\url{10.1109/MM.2018.112130359} (\bibinfo{year}{2018}).

\end{thebibliography}
\end{document}


\section*{Supplementary Material}

\section*{Supplementary note S1: Memory overhead computation}
Continual learning models typically require additional parameters to alleviate catastrophic forgetting. We determine the memory overhead for parameters other than the network weights for different continual learning models \cite{TACOS1, NEO}. Supplementary table \ref{tab:memory_overhead} lists the details of memory overhead computation. The ``Network Configuration'' column shows the number of neurons per layer. ``Network Type'' and ``Weight Precision'' columns list the type of network and precision of weights, respectively. ``MLP'' stands for multi-layer perception and ``SNN'' denotes spiking neural network. The ``CL Parameter Overhead'' column lists the cumulative bit precision of the parameters needed in addition to the weights. These parameters may be required for every weight or neuron, which is denoted in parentheses by W and N, respectively. The ``Normalized Memory Overhead'' column lists the memory overhead for the models considering a network with 784 input neurons, 200 hidden neurons and 2 output neurons. The ``Total Memory Overhead'' column lists the memory overhead of the network used to evaluate the model, as listed in the ``Network Configuration'' column. For models with varying memory overhead, we consider the maximum limit.

\begin{table}[h!]
    \centering
    \caption{Memory overhead computation for different continual learning models.}
    \adjustbox{max width=\textwidth}{
    \begin{tabular}{lcccccc}
        \toprule
         Model & Network Configuration& Network  & Weight  & CL Parameter & Normalized Memory & Total Memory\\
          & & Type &Precision  &Overhead &  Overhead & Overhead\\
       
        \midrule
        LwF \cite{li2017learning}  &  784 - 200 - 200 - 2  & MLP &32-bit &32-bit (W) & 628.8 kB & 788 kB\\
        MAS \cite{Aljundi_MAS}  & 784 - 200 - 200 - 2& MLP &32-bit  & 64-bit (W) & $\sim$ 1.2 MB & 1.6 MB\\
        BGD \cite{zeno2018task}& 784 - 200 - 200 - 2& MLP& 32-bit & 64-bit (W) & $\sim$ 1.2 MB & 1.6 MB\\
         SS \cite{SS} & 784 - 200 - 200 - 2 & MLP & 32-bit & 32-bit (W) & 628.8 kB & 788 kB\\
        TACOS \cite{TACOS1} & 784 - 200 - 200 - 2 & SNN & 32-bit & 48-bit (W) & $\sim$ 0.9 MB & $ 1.2 MB$\\
        NEO \cite{NEO} & 784 - 400 - 400 - 2 & MLP &32-bit&  16-bit (W) & $ \sim$ 0.3 MB & $\sim 1 MB$\\
        NAI \cite{NAI} & 784 - 400 - 400 - 2& MLP &32-bit  & 32-bit (N) & $\sim$ 0.8 kB & 3.2 kB\\
        Activity-dependent Metaplasticity & 784 - 200 - 2& SNN& Multi-memristor&48-bit (W)& $ \sim$ 0.9 MB & $ \sim$ 0.9 MB\\
        w/ Gradient Accumulation & & &weights & & \\
        Probabilistic Metaplasticity & 784 - 200 - 2 & SNN & Multi-memristor & 16-bit (W) &  $ \sim$ 0.3 MB & $ \sim$ 0.3 MB\\
         & &   & weights&16-bit (N) & $\sim$ 0.4 kB & $\sim$ 0.4 kB\\
        \bottomrule
    \end{tabular}}
    \label{tab:memory_overhead}
\end{table}
\begin{table}[h]
    \centering
    \caption{Performance of a spiking network on the split-Fashion MNIST task in different settings. The table shows the individual task accuracies and the mean accuracy across tasks after sequential training. The mean and standard deviation were determined over 5 runs. }
    \adjustbox{max width=\textwidth}{
    \begin{tabular}{lcccccc}
        \toprule
         Task ID & Baseline & Random Consolidation & \multicolumn{3}{c}{Decaying Probabilistic Plasticity} & Probabilistic Metaplasticity \\
        \cmidrule(lr){4-6}
        & &  & (Factor=2) & (Factor=5) & (Factor=10) & \\
        \midrule
        Task 1 & 49.94 $\pm$ 0.06 & 52.84 $\pm$ 2.74 & 50.13 $\pm$ 0.19 & 89.28 $\pm$ 2.80 & 94.07 $\pm$ 0.71 & 95.65 $\pm$ 0.24 \\
        Task 2 & 50.06 $\pm$ 0.15 & 50.21 $\pm$ 0.13 & 50.12 $\pm$ 0.22 & 57.74 $\pm$ 1.30 & 85.51 $\pm$ 2.04 &  79.02 $\pm$ 1.09\\
        Task 3 & 88.22 $\pm$ 5.00 & 96.08 $\pm$ 0.67 & 97.75 $\pm$ 0.53 & 98.03 $\pm$ 0.58 & 87.44 $\pm$ 3.11 & 95.99 $\pm$ 1.42 \\
        Task 4 & 86.86 $\pm$ 9.84 & 98.69 $\pm$ 0.95 & 99.43 $\pm$ 0.30 & 99.92 $\pm$ 0.05  & 97.76 $\pm$ 0.67 &  98.77 $\pm$ 0.11\\
        Task 5 & 99.67 $\pm$ 0.05 & 99.61 $\pm$ 0.07 & 99.38 $\pm$ 0.06  & 91.75 $\pm$ 1.32 & 72.61 $\pm$ 5.80 & 96.72 $\pm$ 0.35 \\
        \midrule
        Mean & 74.95 $\pm$ 2.87  & 79.48 $\pm$ 0.76 & 79.36 $\pm$ 0.24 & 87.34 $\pm$ 0.88 & 87.48 $\pm$ 1.61 & 93.23 $\pm$ 0.14 \\
        \bottomrule
    \end{tabular}}
    \label{tab:control_exp_fmnist}
\end{table}
\section*{Supplementary note S2: Controlled experiments on the split-Fashion MNIST task}
Supplementary table \ref{tab:control_exp_fmnist} shows the results of the controlled experiments on the split-Fashion MNIST task.  In random consolidation, the update probabilities of the weights are calculated as probabilistic metaplasticity, but the results are randomly shuffled among the weights. In decaying probabilistic plasticity, all weights are updated with the same probability and the update probability is reduced by a factor with each incoming task to help retain previous knowledge.
 All experiments were carried out considering a spiking network with a single hidden layer of 200 neurons and 7 memristor weights.
The baseline network is trained with eRBP combined with error threshold. From Table \ref{tab:control_exp_fmnist}, we see that random consolidation shows performance similar to the baseline with catastrophic forgetting of the initial two tasks. Decaying probabilistic plasticity either remembers old tasks well or learns new tasks well, depending on the decaying plasticity factor with a poor balance of plasticity and stability. Probabilistic metaplasticity shows superior balance as it remembers initial tasks while being able to learn the final task with lower accuracy degradation compared to decaying probabilistic plasticity.

\begin{algorithm}[t]
\DontPrintSemicolon
\SetAlgorithmName{Supplementary Algorithm}
\SetKwComment{Comment}{$\triangleright$\  }{}
\SetKwComment{Commente}{$\#$ }{}

\SetKwFunction{Weight}{WeightUpdate}

    \For{$task \in \mathcal{T}$}{

   \For{$\{x^t,y^t\} \in \{X^t,Y^t\}$}{
   
   \For{\var{t} $\in$ $T_{\text{sim}}$}{
   Forward Pass: $S^{\text{out}}(t)$ $\gets$ $f (S^{\text{in}}(t), W(t)$\; Random Error Feedback: $\tau_U \frac{\partial \mathrm{U}}{\partial t} =  E\mathrm{R}_U$\; 
   Update Neuron Trace: $\frac{d}{dt}X_{\text{tr}} = -\frac{X_{\text{tr}}}{\tau_{\text{tr}}} +S$ \;  
 
  \For {\var{i} $\in$ $\{X_i(t)==1\}$}{
   \If{$I_{\text{min}}<I_j<I_{\text{max}}$}{
   $U_{\text{syn},ij} \gets -\eta U_j f(m_{ij},w_{ij})$ \;
   }
   }

  \If{$|U_{\text{syn},ij}|>U_{\text{th}}$}{
   Update memristor weights: $W$ $\gets$ $Write (R_{\text{mem}}, U)$\;
   $U_{\text{syn},ij}$ = 0\;
   
  }
}
  Update metaplasticity coefficient: $m \gets m + {\Delta}m$ \;
   }
   }
\caption{Activity-dependent metaplasticity with gradient accumulation for continual learning. $\mathcal{T}$ denotes a set of sequential tasks. $S^{\text{in}}$ and $S^{\text{out}}$ are the input and output spike trains of the network and $W$ denotes the memristor weights. $U$ is the error accumulated at the dendritic compartments. As the network trains, the gradients of the weights eligible for update are accumulated in high-precision memory $U_{\text{syn}}$. When the accumulated gradient crosses a threshold $U_{\text{th}}$, the memristor weights are programmed to the next higher or lower conductance level depending on the sign of the accumulated error. The function Program() refers to the operations required to update memristor conductance based on the error.}
\label{alg:acc}
\end{algorithm}

\section*{Supplementary note S3: Energy analysis}

Supplementary table \ref{tab:en_analysis_params} lists the design parameters for the energy analysis in the mixed-signal design. We designed a $16\times16$ crossbar with programming peripherals in the 65nm technology node with 1.2 V and 3.3 V supply voltage. The row and column peripherals consist of transmission gate-based multiplexers and level shifters to supply the appropriate voltages to the crossbar terminals required during different network operations. The memristor weights are implemented with a Verilog-A model that captures the behavior of the 1T1R memristor device considered in this work \cite{verilogA_model}. Operational amplifiers and voltage comparators were used for the read and boxcar function \cite{opamp_for_weight}.
The modules needed for the operations carried out in the digital domain were synthesized in the IBM 65nm technology node considering 1.2 V supply and 100 MHz frequency in Synopsys Design Compiler. \\
\begin{table}[h]
    \centering
    \caption{Mixed-signal design parameters for energy analysis. }
    \begin{tabular}{lc}
        \toprule
         Parameter & Value  \\
        \\
        \midrule
        Technology node & 65 nm  \\
        Supply voltage & 1.2 V, 3.3 V\\
        Frequency & 100 MHz  \\
        Memristor devices per weight & 7 \\
        Memristor resistance range & 3.5 $k \Omega$ - 25 $k\Omega$ \\
        Set/Reset/Read voltage & 2/1.6/0.1 V \\
        ADC resolution & 8-bit \\
        \bottomrule
    \end{tabular}
    \label{tab:en_analysis_params}
\end{table}

\begin{table}[h]
    \centering
    \caption{Energy for network operations. }
    \begin{tabular}{lc}
        \toprule
         Operation & Energy Consumption (pJ)  \\
        \\
        \midrule
        Memristor weight update & 38.6 \\
        Memristor weight read & 23.8 \\
        Addition (16-bit/32-bit)& 0.26/0.6 \\
        Multiplication & \\
        (16, 8 / 16, 16/ 16, 32)& 0.98/1.98/4 \\
        Shift operation & 0.2 \\
        Comparison (16-bit / 32-bit) & 0.24/0.48 \\
        Random number generation & 0.42\\
        Exponential metaplasticity function & 7.962 \\
        0.6 MB SRAM 32-bit read/write  & 144.3/142.5 \\
        0.3 MB SRAM 16-bit read/write  & 124.1/103.7  \\
        39.4 kB SRAM 16-bit read/write  & 68.76/37.25  \\
        0.5 kB SRAM 16-bit read/write  & 5.4/6.3 \\
        Register access & 0.04795 \\
        \bottomrule
    \end{tabular}
    \label{tab:en_analysis_val}
\end{table}
\noindent
\textbf{Probabilistic Metaplasticity Approach:}
When training with probabilistic metaplasticity, the accumulated error is compared with the error threshold at each timestep during the spike train presentation. This operation was carried out with 16-bit digital comparator. If the error is above the threshold, the currents at the post-synpatic neurons are converted to voltage with an operational amplifier, which is then propagated to voltage comparators to evaluate the boxcar function. The energy for this operation is estimated considering the average number of active spikes in the network and assuming the memristor resistance to be the average of the resistance levels used to map weights. We consider a read voltage of 0.1 V with 100 ns duration. The average power consumption of the operational amplifier is $\sim$ 24 $\mu$W, with 1.2 V supply and $\sim$ 20 $\mu$A average current.  The reference for the operational amplifier consumes about 13.5 $\mu$W. The output of the operational amplifier is connected to the boxcar function unit, which consists of two voltage comparators and an AND gate. The energy is estimated as -

$E_{\text{cc}}$ = ($\frac{V^2_{\text{read}}}{R_{\text{weight,avg}}} + 2 \times  P_{\text{opamp}} + P_{\text{reference}} +P_{\text{boxcar unit}}) \times t_{\text{read}}$

\vspace{6pt}
For the weights eligible for update, the update probability is computed and compared with a random number. If the random number is lower than the update probability of a memristor, it is updated. We designed a 32-bit Linear Feedback Shift Register (LFSR) to generate the random number. Since the update probability depends on the weight magnitude, this requires a read weight operation. During this operation, the columns of the crossbar are connected to an operational amplifier, the output of which can be read by an Analog to Digital Converter (ADC). We estimate the energy consumed by the memristor device assuming it to be the average of resistance levels used to map the weight. The energy for 8-bit ADC in 65nm technology node is estimated according to \cite{adc_energy} to be 19.46 pJ. This leads to read weight energy determined as -\\

$E_{\text{read weight}}$ = $(P_{\text{opamp}} + P_{\text{reference}} + \frac{V^2_{\text{read}}\times n_{\text{mem}}}{R_{\text{mem,avg}}}) \times t_{\text{read}} + E_{\text{ADC}}$

\vspace{6pt}
The metaplasticity coefficients require 16-bit precision and are emulated with on-chip SRAM. The energy required to compute the exponential metaplasticity function was determined with a pipelined exponential function module \cite{exponent_function}. The output of the function is compared with the random number generated with LFSR to determine which weights need to be updated.

The weight update involves programming the 1T1R devices, which can be set to different conductance levels by controlling the compliance current with the gate voltage in the transistor \cite{liehr2020impact}. Each weight update operation consists of a read operation to determine the current conductance level, then a reset and set operation to program it to the next higher or lower conductance level based on the sign of the error. The read energy is estimated with the same considerations as the read weight operation. During set (reset), the selected row and column are set to 2 V (0 V) and 0 V (1.6 V), respectively, for 100 ns \cite{jubin_switchtime}. The set (reset) energy varies with the target (current) conductance level of the device. We determine the energy consumption for all conductance levels and use the average energy value for the analysis.\\
$E_{\text{read}}$ = $(P_{\text{opamp}} + P_{\text{reference}} + \frac{V^2_{\text{read}}}{R_{\text{mem,avg}}}) \times t_{\text{read}} + E_{\text{ADC}}$\\
$E_{\text{set}}$ = $P_{\text{set,avg}} \times t_{\text{set}}$ \\
$E_{\text{reset}}$ = $P_{\text{reset,avg}} \times t_{\text{reset}}$  \\
$E_{\text{weight program}}$ = $E_{\text{read}}$ + $E_{\text{set}}$ + $E_{\text{reset}}$ \\

\vspace{6pt}
The metaplasticity coefficients are emulated as on-chip SRAM with 16-bit bus width. We determine the SRAM read and write energy with HP Cacti \cite{muralimanohar2009cacti}. For individual, module-shared and neuron-shared metaplasticity coefficients, the on-chip SRAM block was instantiated  as 0.3 MB, 39.4 kB and 0.5 kB respectively. The layer-shared metaplasticity coefficients were realized with registers.\\

\noindent
\textbf{Activity-dependent metaplasticity with gradient accumulation:}
The gradient accumulation approach computes the gradients of the weights eligible for update according to eRBP. Checking the update eligibility requires evaluating the boxcar function and computing the gradient requires evaluating the exponential metaplasticity function. The energy for both these operations are determined in the same manner as the probabilistic approach. The output of the metaplasticity function (32-bit) is multiplied by the dendritic compartment error (16-bit), and the learning rate is modeled with a diadic number which replaces the multiplication with a shift operation. The calculated gradient is compared with the threshold to determine the weights to be updated and then accumulated in 32-bit on-chip SRAM. 
The on-chip SRAM required for the gradients and the metaplastic coefficients were modeled as 0.6 MB block with 32-bit bus width and 0.3 MB block with 16-bit bus width, respectively. 

The energy estimates for the different operations for the analysis are listed in Supplementary table \ref{tab:en_analysis_val}. 

\begin{table}[h]
    \caption{Performance of a spiking network with probabilistic metaplasticity with individual and shared metaplasticity coefficients on the split-MNIST task. The table shows the mean and standard deviation of the individual task accuracies and the mean accuracy across tasks over 5 runs after sequential training.}
\adjustbox{max width=\textwidth}{
    \begin{tabular}{lcccc}
        \toprule
          \#Task& Individual Metaplasticity  & Module-shared Metaplasticity & Neuron-shared Metaplasticity & Layer-shared Metaplasticity\\
         & Coefficients  & Coefficients & Coefficients & Coefficients \\
         
        \midrule
        Task 1 &84.95 $\pm$ 3.42 & 88.77 $\pm$ 4.21 & 90.00 $\pm$ 3.65& 89.49 $\pm$ 7.37\\
        Task 2 & 90.20 $\pm$ 1.96&86.71 $\pm$ 4.14&90.04 $\pm$ 2.33 & 88.49 $\pm$ 5.78\\
        Task 3 & 72.06 $\pm$ 2.60& 68.67 $\pm$ 3.99& 70.11 $\pm$ 4.27&69.16 $\pm$ 5.83\\
        Task 4 & 94.73 $\pm$ 0.94&93.21 $\pm$ 0.74 & 91.53 $\pm$ 2.99 & 89.06 $\pm$ 1.39\\
        Task 5 & 76.54 $\pm$ 2.60&70.16 $\pm$ 4.77 & 68.68 $\pm$ 4.64&62.49 $\pm$ 3.34\\
        \midrule
        Mean & 83.70 $\pm$ 0.78& 81.50 $\pm$ 1.18& 82.07 $\pm$ 0.84 & 79.74 $\pm$ 1.21 \\
        \bottomrule
    \end{tabular}}%

    \label{tab:shared_m_splitmnist}

\end{table}

\begin{table}[h]
    \centering
    \caption{Classification accuracies of a spiking network with multi-memristor weights ($n_{mem}$ = 7) on the split-MNIST and split-Fashion MNIST tasks. }
    \adjustbox{max width=\textwidth}{
    \begin{tabular}{lcccc}
        \toprule
          & \multicolumn{2}{c}{Split-MNIST  } &\multicolumn{2}{c}{Split-Fashion MNIST}  \\
           \cmidrule(lr){2-3}
           \cmidrule(lr){4-5}
           
          \# Task & Gradient Accumulation & Error-threshold training & Gradient Accumulation & Error-threshold training \\
           \cmidrule(lr){1-5}
           
           Task 1 & 99.91 $\pm$ 0.07 & 99.93 $\pm$ 0.02 & 97.43 $\pm$ 0.66& 96.35 $\pm$ 1.58\\
           Task 2& 96.52 $\pm$ 0.86  & 96.56 $\pm$ 0.91 & 96.07 $\pm$ 0.95 & 96.28 $\pm$ 0.51\\
           Task 3& 98.22 $\pm$ 0.45 &  98.93 $\pm$ 0.18& 99.84 $\pm$ 0.12 & 99.94 $\pm$ 0.04\\
           Task 4& 99.58 $\pm$ 0.14  & 99.51 $\pm$ 0.09 & 99.93 $\pm$ 0.04& 99.97 $\pm$ 0.02\\
           Task 5&  96.31 $\pm$ 1.01 & 97.57 $\pm$ 0.31& 99.61 $\pm$ 0.18  & 99.69 $\pm$ 0.04\\

        \bottomrule
    \end{tabular}}
    \label{tab:supp_class_experiment}
\end{table}